\documentclass[aps,prl,reprint,superscriptaddress]{revtex4-2}

\usepackage{graphicx}

\begin{document}

\title{Exotic magnetic phase diagram and extremely robust antiferromagnetism in Ce$_2$RhIn$_8$}

\author{T.~Klein}
\affiliation{Universit\'{e} Grenoble Alpes, CNRS, Institut N\'{e}el, 38000 Grenoble, France}

\author{C.~Marcenat}
\affiliation{Universit\'{e} Grenoble Alpes, CEA, Grenoble-INP, IRIG, Pheliqs, 38000 Grenoble, France}

\author{A.~Demuer}
\affiliation{Laboratoire National des Champs Magn\'{e}tiques Intenses (LNCMI-EMFL), CNRS, Universit\'{e} Grenoble Alpes, 38042 Grenoble, France}

\author{J.~Sarrade}
\affiliation{Laboratoire National des Champs Magn\'{e}tiques Intenses (LNCMI-EMFL), CNRS, UUniversit\'{e} Grenoble Alpes, 38042 Grenoble, France}

\author{D.~Aoki}
\affiliation{IMR, Tohoku University, Oarai, Ibaraki 311-1313, Japan}

\author{I. Sheikin}
\email[]{ilya.sheikin@lncmi.cnrs.fr}
\affiliation{Laboratoire National des Champs Magn\'{e}tiques Intenses (LNCMI-EMFL), CNRS, Universit\'{e} Grenoble Alpes, 38042 Grenoble, France}

\date{\today}

\begin{abstract}
The antiferromagnetic heavy-fermion compound Ce$_2$RhIn$_8$ belongs to the same family and bears similarities with the well-studied prototypical material CeRhIn$_5$, which demonstrates a unique behavior under applied magnetic field. Here, we report specific-heat measurements on a high-quality single crystal of Ce$_2$RhIn$_8$ in magnetic fields up to 35~T applied along both principal crystallographic directions. When the magnetic field is applied along the $a$ axis of the tetragonal crystal structure, two additional field-induced antiferromagnetic phases are observed, in agreement with previous reports. One of them is confined in a small area of the field-temperature phase diagram. The other one, which develops above $\sim$2.5~T, is very robust against the field. Its transition temperature increases with field, reaches a maximum at $\sim$12~T, and then starts to decrease. However, it tends to saturate towards the highest field of our measurements. For a field applied along the $c$ axis, the N\'{e}el temperature, $T_N$, initially decreases with field, as expected for a typical antiferromagnet. Surprisingly, an additional phase emerges above $\sim$10~T. It is most likely to be of the same origin as its counterpart observed above 2.5~T for the other field orientation. This phase is also unusually robust: Its transition temperature increases all the way up to 35~T, where it exceeds the zero field $T_N =$~2.85~K. Finally, for both field directions, the phase diagrams contain rarely observed tricritical points of second-order phase transitions.
\end{abstract}

\maketitle



Cerium-based intermetallic compounds exhibit a variety of fascinating phenomena including heavy-fermion behavior, quantum criticality, non-Fermi-liquid behavior, and novel states of matter such as unconventional superconductivity~\cite{Pfleiderer2009} and, probably, electronic nematicity~\cite{Ronning2017}. Of particular interest is the Ce$_n$T$_m$In$_{3n+2m}$ ($T =$ transition metal) family, in which $n$ conducting CeIn$_3$ layers are separated by $m$ insulating $T$In$_2$ layers along the $c$ axis. Within this family the dimensionality is reduced with increasing $m/n$ ratio from cubic and fully three-dimensional CeIn$_3$ ($n = 1$, $m = 0$) to the most two-dimensional CePt$_2$In$_7$ ($n = 1$, $m = 2$).

The ground state of some of the family members, such as CeIn$_3$, CeRhIn$_5$ ($n = 1$, $m = 1$), and CePt$_2$In$_7$, is antiferromagnetic (AFM). All these compounds can be tuned to a magnetic quantum critical point (QCP), i.e., a continuous magnetic phase transition at zero temperature, by hydrostatic pressure $P$ or magnetic field $H$. When tuned by pressure, unconventional superconductivity emerges in the vicinity of the QCP~\cite{Mathur1998,Hegger2000,Bauer2010a}. The critical fields $H_c$ for the suppression of the AFM order in all three materials are very high, i.e., 60\---80~T for CeIn$_3$~\cite{Ebihara2004,Moll2017}, $\sim$50~T for CeRhIn$_5$~\cite{Jiao2015,Jiao2019}, and $\sim$60~T for CePt$_2$In$_7$~\cite{Krupko2016}. Such high magnetic fields can be generated by pulsed magnets only, which limits both the number of available experimental techniques and the temperature range of possible experiments. That is why the presence of field-induced QCPs in these materials has yet to be experimentally confirmed beyond any doubt, and, therefore, strictly speaking, remains a hypothesis. For instance, it has yet to be checked that the AFM transitions remain second order down to sufficiently low temperatures when approaching putative QCPs. Furthermore, the non-Fermi-liquid behavior, one of the fingerprints of a QCP, has not been experimentally observed either.

For anisotropic tetragonal compounds, CeRhIn$_5$ and CePt$_2$In$_7$, $T_N$ monotonically decreases with field applied along the $c$ axis, while it first increases at low fields and then starts to decrease when the field is applied in the basal plane~\cite{Krupko2016,Mishra2021}. Furthermore, in CeRhIn$_5$, an additional phase transition was observed at about 30~T applied along (or close to) the $c$ axis within the AFM state~\cite{Moll2015,Jiao2015,Ronning2017,Rosa2019,Kurihara2020,Lesseux2020,Mishra2021,Mishra2021a}. The exact origin of this transition is still under debate. In addition, two more field-induced phases were observed for $B \parallel a$~\cite{Cornelius2001,Mishra2021}.

Ce$_3$PdIn$_{11}$ and Ce$_3$PtIn$_{11}$ belonging to the same family ($n = 3$, $m = 1$) exhibit two successive AFM transitions followed by the emergence of bulk superconductivity already at ambient pressure~\cite{Kratochvilova2015a,Prokleska2015}. Their magnetic phase diagrams established in moderate magnetic fields up to 9~T are even more complex. For $H \parallel c$, the two AFM transition temperatures first decrease with field, then merge together, and then split again~\cite{Das2019,Das2018}. In addition, a field-induced first-order phase transition develops at low temperatures in both compounds.

In order to get further insight into field-induced QCPs, it is important to search for other compounds of the family, in which a QCP can be induced at a lower field accessible in static-field magnets. It is equally important to establish complete magnetic phase diagrams of the other AFM members of the family in order to better understand the origin of the field-induced phases in these compounds.

Ce$_2$RhIn$_8$ is another member of the family discussed above with $n = 2$ and $m = 1$. It crystallizes into a tetragonal crystal structure with space group $P4/mmm$ (No. 123)~\cite{Grin’1979}. The electronic specific-heat coefficient $\gamma \sim$~400~mJ/molK$^2$~\cite{Cornelius2001} characterizes Ce$_2$RhIn$_8$ as a moderate heavy-fermion compound. The material undergoes an AFM transition at a N\'{e}el temperature $T_N \simeq$ 2.8~K~\cite{Thompson2001}. The magnetic structure determined from single-crystal neutron diffraction is commensurate with the propagation vector $\mathbf{Q} = (1/2, 1/2, 0)$ and the staggered moment of 0.55$\mu_B$~\cite{Bao2001}. What is very unusual here is that magnetic moments are tilted by 38$^\circ$ from the $c$ axis towards the basal plane~\cite{Bao2001}. In other Ce-based antiferromagnets with a tetragonal crystal structure, magnetic moments are either aligned in the basal plane or along the $c$ axis.



Previous specific-heat measurements~\cite{Cornelius2001} performed in magnetic fields up to 10~T revealed a monotonic decrease of $T_N$ for $H \parallel c$. A naive extrapolation of $T_N(H)$ to zero suggests a field-induced QCP at about 30~T. On the other hand, for $H \parallel a$, two additional field-induced phases of currently unknown origin were observed. All the phase transitions are found to be second order.

The above results suggest that specific-heat measurements in higher fields are required in order to verify the presence of a QCP for $H \parallel c$ and to complete the phase diagram for fields along the $a$ axis.

In this Letter, we report high-field low-temperature specific-heat measurements on a single crystal of Ce$_2$RhIn$_8$. The measurements were performed in static fields up to 35~T for field orientations both along the $a$ and $c$ axes. For a field applied along the $a$ axis, we observed all the previously reported transitions and traced one of them to much higher fields. For a field along the $c$ axis, we observed another phase, which develops above $\sim$10~T at low temperature. This phase becomes dominant at high fields and its transition temperature increases all the way up to 35~T where it exceeds the zero field $T_N$. Based on these results, we propose a revision of the magnetic phase diagram of Ce$_2$RhIn$_8$ for both principal orientations of the magnetic field.

\begin{figure}[htb]
\includegraphics[width=\columnwidth]{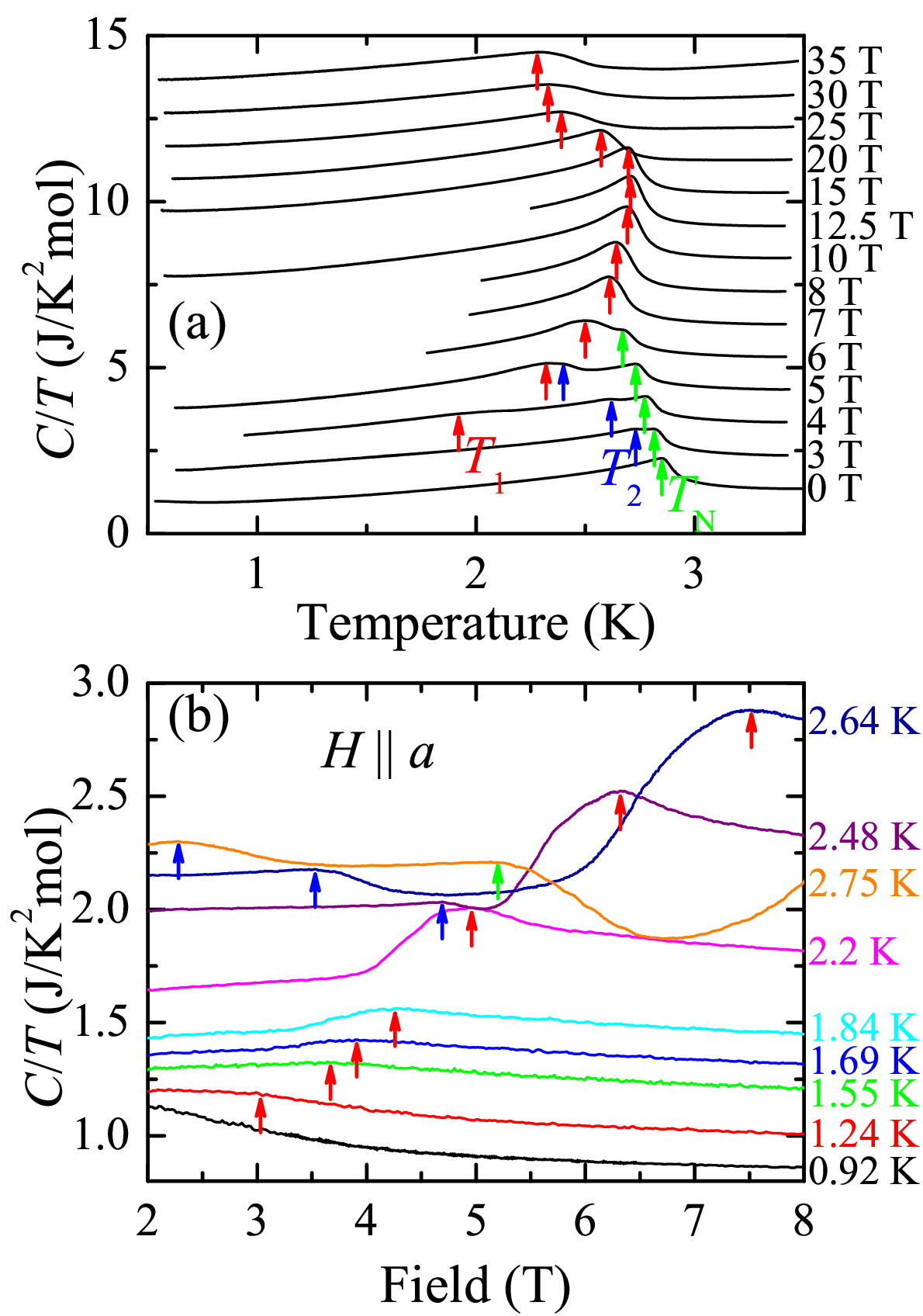}
\caption{\label{fig:Sweeps_a} Specific heat divided by temperature, $C/T$,  of Ce$_2$RhIn$_8$ for a magnetic field applied along the $a$ axis. (a) $C/T$ as a function of $T$ for several values of the magnetic field. Curves are vertically shifted for clarity. (b) $C/T$ as a function of $H$ measured at different temperatures. Arrows indicate anomalies corresponding to phase transitions, as discussed in the text.}
\end{figure}


A high-quality platelet-like single crystal with a mass of (24$\pm$1)~$\mu$g used in the present study was grown by the In self-flux method. Details of the sample preparation are given elsewhere~\cite{Ueda2004}. The platelet surface is perpendicular to the crystallographic $c$ axis. Specific-heat measurements were performed by an ac technique~\cite{Kraftmakher2002} in a $^3$He refrigerator down to 600~mK and up to 35~T in the M9 resistive magnet of the LNCMI-Grenoble. Additional low-field measurements were performed in a $^4$He flow cryostat in an 8~T superconducting magnet. Details of the specific-heat setup are given elsewhere~\cite{Michon2019}.


\begin{figure}[htb]
\includegraphics[width=\columnwidth]{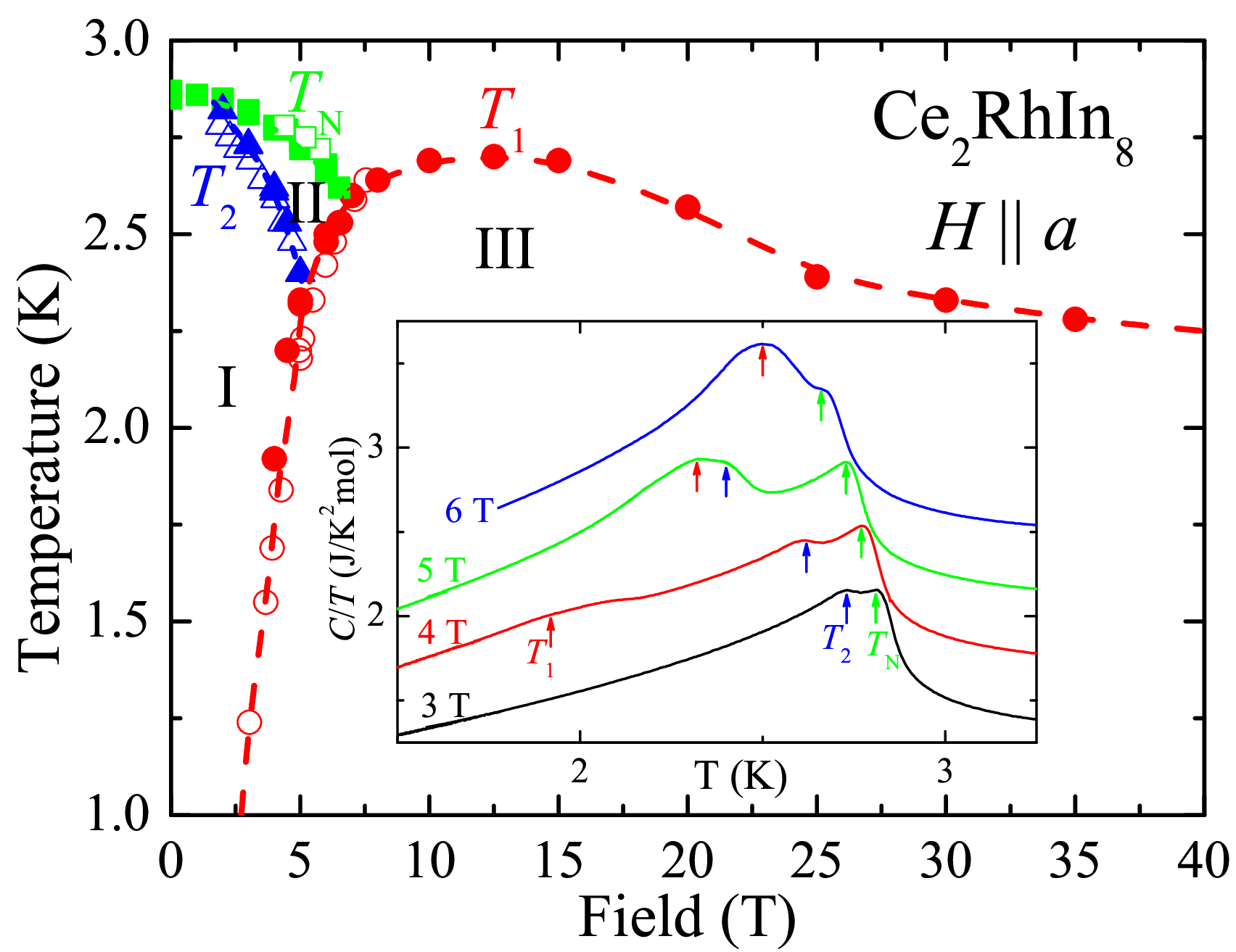}
\caption{\label{fig:Phas_diag_a} Magnetic phase diagram of Ce$_2$RhIn$_8$ obtained from specific-heat measurements performed as a function of temperature (solid symbols) and magnetic field (open symbols) for a field applied along the $a$ axis. Lines are a guide for the eye. The inset shows a zoom on the low-field temperature sweeps, for which multiple anomalies appear on the $C/T$ vs $T$ curves. Curves are vertically shifted for clarity.}
\end{figure}

Figure~\ref{fig:Sweeps_a}(a) shows the temperature dependence of the specific heat divided by temperature, $C/T$, at different magnetic fields applied along the $a$ axis. For this field orientation, apart from the AFM transition at $T_N$, there are two additional field-induced transitions at $T_1$ and $T_2$, as shown in more detail in the inset of Fig.~\ref{fig:Phas_diag_a}. We have not observed any hysteresis between up and down temperature sweeps. Furthermore, we have not observed any sharp $\delta$-like peaks characteristic of a first-order transition. Thus, all the transitions are second order. The same conclusion was drawn from previous lower-field specific heat measurements~\cite{Cornelius2001}. In agreement with a previous report~\cite{Cornelius2001}, $T_N$ decreases with magnetic field. It was observed up to 6~T, above which it gives way to the transition at $T_1$. The latter first increases up to about 12~T, and then decreases all the way up to the highest field of our measurements. Remarkably, $T_1(H)$ flattens and almost saturates towards 35~T, the highest field of our measurements. Finally, the transition at $T_2$ is observed only over a very limited field range from 3 to 5~T.

All the transitions are observed also in magnetic field sweeps performed at constant temperatures, as shown in Fig.~\ref{fig:Sweeps_a}(b). At higher temperatures, the $T_1$ transition manifests itself as a clear maximum in $C/T$ vs $H$. This feature becomes smaller and smaller with decreasing temperature, changes to a kink at 1.25~K, and is no longer observed at 0.9~K.

The resulting magnetic field temperature, $H\--T$, phase diagram for field along the $a$ axis is shown in Fig.~\ref{fig:Phas_diag_a}. As was already mentioned above, the zero-field phase I is commensurate AFM with a propagation vector $\mathbf{Q} = (1/2, 1/2, 0)$. The exact origin of the two other field-induced phases, II and III, is unknown at present. Nevertheless, both phases are certainly AFM. Indeed, the phase II emerges directly below $T_N$ at low fields. As for phase III, previous neutron-diffraction experiments, although performed for $H \parallel [110]$, revealed a finite intensity of the magnetic Bragg peak (1/2, 1/2, 1) up to 14.5~T at low temperatures~\cite{Moshopoulou2002}.

The magnetic phase diagram can be analyzed using the magnetic Ehrenfest equation applicable to a second-order phase transition,

\begin{equation}\label{Ehrenfest}
\Delta\left( \frac{C}{T} \right)_H = -\Delta\left( \frac{\partial M}{\partial T} \right)_H \frac{d H}{d T},
\end{equation}
where $\Delta (C/T)$ is the size of the specific-heat jump at the phase transition, and $M$ is the magnetization. First, let us note that at low field $dH/dT$ at $T_1$ decreases with decreasing temperature, which accounts for the shrinking of the specific-heat anomaly at the phase boundary. The anomaly disappears when $dH/dT \rightarrow 0$, i.e., when the phase boundary becomes almost temperature independent. Similarly, $dH/dT$ at $T_1$ increases towards the maximum of $T_1$ at about 12~T. Accordingly, the size of the specific-heat jump grows in the vicinity of this field, as can be seen in Fig.~\ref{fig:Sweeps_a}(a). At the maximum of $T_1(H)$, $dH/dT \rightarrow \infty$, but $\Delta (C/T)$ remains finite. This implies that $\Delta (\partial M / \partial T) \rightarrow 0$, i.e., there should be no anomaly on the $M(T)$ curve at about 12~T. Below this field, both $\Delta (C/T)$ and $dH/dT$ are positive. Therefore, $\Delta (\partial M / \partial T)$ is negative, i.e., the slope of the $M(T)$ curve should decrease when crossing the phase transition at $T_1$. Above 12~T, $dH/dT < 0$ while $\Delta (C/T) > 0$, meaning that the slope of the $M(T)$ curve should increase. It would be interesting to perform temperature-dependent magnetization measurements at different fields to experimentally confirm this conclusion.

\begin{figure}[htb]
\includegraphics[width=\columnwidth]{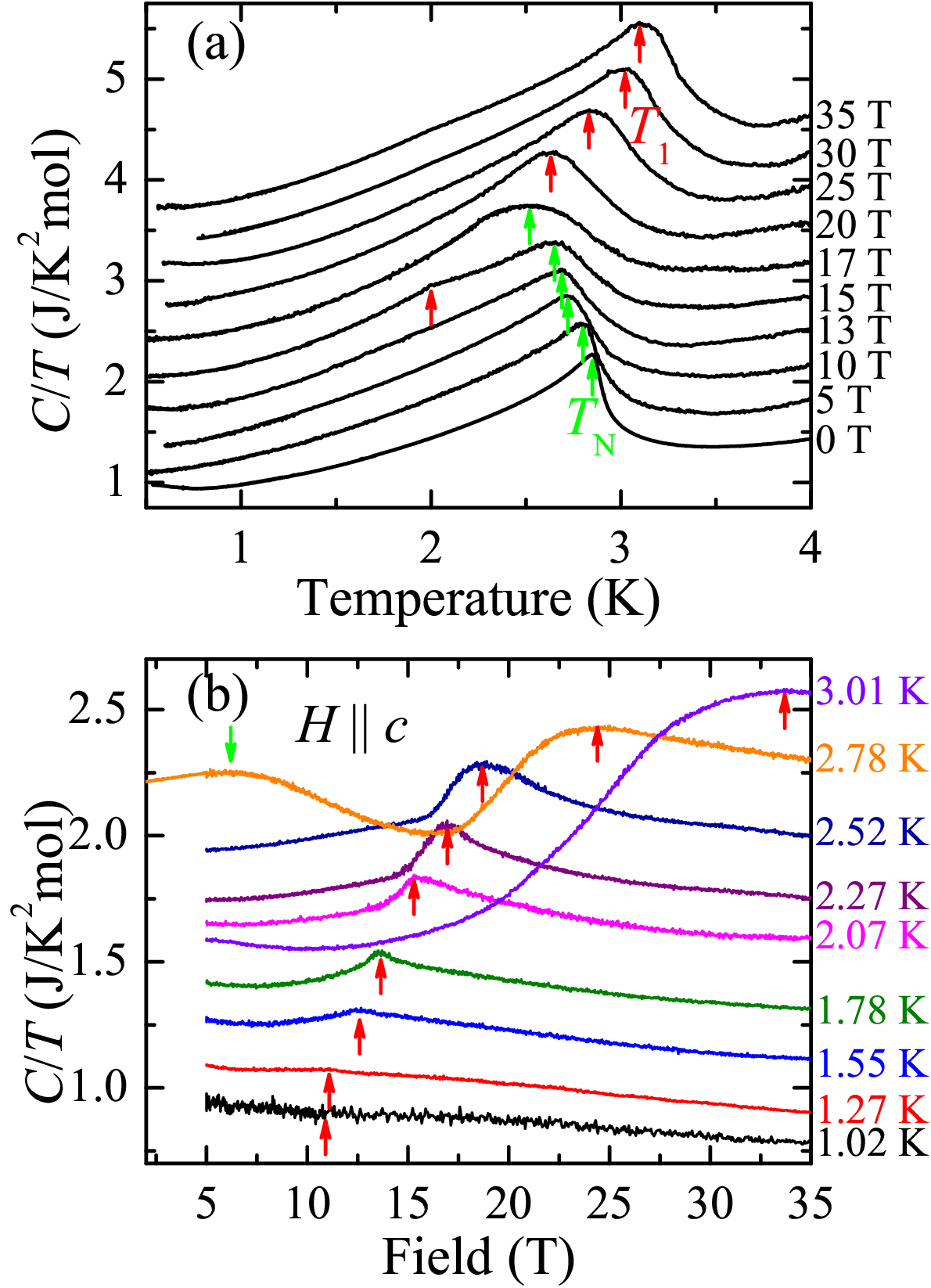}
\caption{\label{fig:Sweeps_c} $C/T$  of Ce$_2$RhIn$_8$ for $H \parallel c$. (a) $C/T$ vs $T$ for several values of the magnetic field. Curves are vertically shifted for clarity. (b) $C/T$ vs $H$ measured at different temperatures. Arrows indicate anomalies corresponding to phase transitions.}
\end{figure}

We will now turn our attention to the results obtained for $H \parallel c$. Figure~\ref{fig:Sweeps_c}(a) shows $C/T$ as a function of $T$ at different magnetic fields applied along the $c$ axis. For this orientation of the magnetic field, $T_N$ is gradually suppressed up to 17~T, above which it is no longer observed. The most surprising result is the emergence of another transition at $T_1$ above 15~T. This transition is also second order. The transition temperature $T_1$ increases with field all the way up to 35~T and exceeds the zero field $T_N =$~2.85~K above $\sim$25~T.

The transition at $T_1$ also manifests itself as an anomaly in $C/T$ in field sweeps, as shown in Fig.~\ref{fig:Sweeps_c}(b). Similar to the other orientation, the size of the anomaly decreases with decreasing temperature and finally becomes undetectable below $\sim$1~K. This is because the phase boundary becomes almost temperature independent at low temperatures. This leads to vanishing of the specific-heat jump $\Delta (C/T)$ according to Eq.~\ref{Ehrenfest}.

\begin{figure}[htb]
\includegraphics[width=\columnwidth]{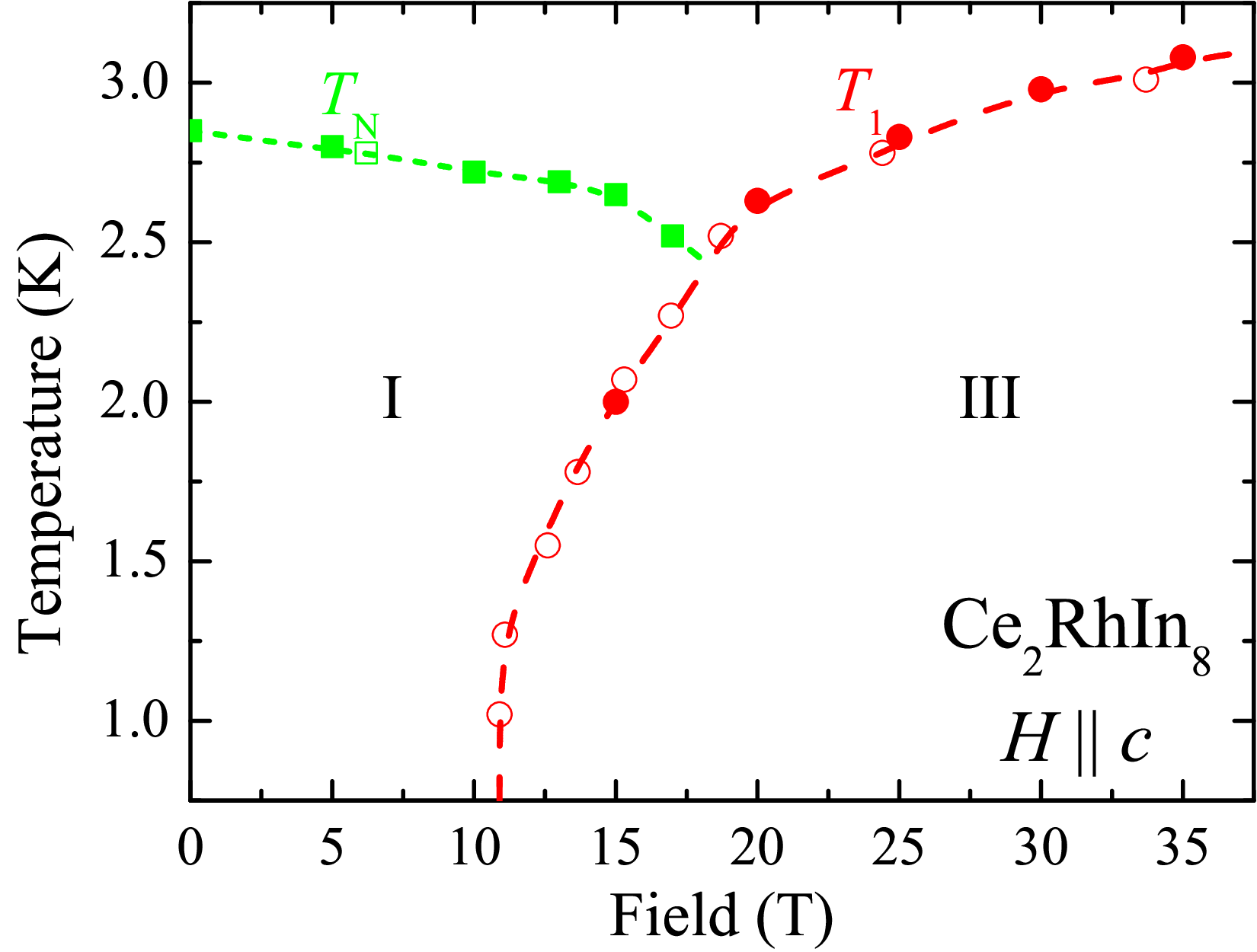}
\caption{\label{fig:Phas_diag_c} Magnetic phase diagram of Ce$_2$RhIn$_8$ obtained from $T$ sweeps (solid symbols) and $H$ sweeps (open symbols) for $H \parallel c$. Lines are a guide for the eye.}
\end{figure}

Figure~\ref{fig:Phas_diag_c} shows the revised magnetic phase diagram of Ce$_2$RhIn$_8$ for a field applied along the $c$ axis. Here, also the origin of the field-induced phase III is unknown, but it is likely to have the same or similar magnetic structure as its counterpart for $H \parallel a$. Indeed, its phase boundary is qualitatively similar to that for $H \parallel a$ below 12~T. Assuming naively that the field value of the maximum of $T_1$ scales with the onset field of phase III, a maximum of $T_1$ is expected at $\sim$50~T for $H \parallel c$.

For both orientations of the magnetic field, all the phase transitions are second order. For $H \parallel c$, there is a tricritical point where $T_N$ merges with $T_1$. For $H \parallel a$, the phase diagram (Fig.~\ref{fig:Phas_diag_a}) contains three such points where three boundaries of second-order phase transitions meet together. Thermodynamically, such a situations is possible only if two transition lines with finite specific-heat jumps $\Delta C$ have equal slopes, whereas the third boundary exhibits the specific-heat anomaly that vanishes towards the tricritical point~\cite{Braithwaite2019}. This condition is satisfied, at least qualitatively, in Ce$_2$RhIn$_8$. Indeed, for $H \parallel c$, there is no kink in the $T_1$ phase boundary in the vicinity of the tricritical point, and the specific-heat jump decreases upon approaching it [see Fig.~\ref{fig:Sweeps_c}(a)]. A similar trend is observed for $H \parallel a$ at the tricritical points where $T_N$ and $T_2$ merge with $T_1$. It is difficult to estimate quantitatively the size of the specific-heat jumps upon approaching the tricritical points due to the presence of two specific-heat anomalies very close to each other. Tricritical points of second-order phase transitions, commonly believed to be thermodynamically forbidden~\cite{Yip1991}, are rare. To the best of our knowledge, the spin-triplet superconductor UTe$_2$ is the only other system, in which similar crossings of three lines of second-order transitions were recently observed~\cite{Braithwaite2019,Rosuel2023}.

Magnetic phase diagrams of Ce$_2$RhIn$_8$ established here bear certain similarities with those of its sister compound CeRhIn$_5$, in which an incommensurate AFM order emerges below $T_N =$~3.8~K~\cite{Hegger2000,Bao2000}. In CeRhIn$_5$, a magnetic field applied in the basal plane also induces two additional phase transitions~\cite{Cornelius2001,Correa2005,Mishra2021}. The higher-temperature transition is also second order. It corresponds to a change of the ordered moment, while the propagation vector, almost the same as in a zero magnetic field~\cite{Raymond2007}, becomes temperature dependent~\cite{Fobes2018}. This transition is also observed at low fields only. It is possible that the transition at $T_2$ observed in Ce$_2$RhIn$_8$ is of the same or similar origin. However, the lower-temperature transition observed in CeRhIn$_5$ for $H \parallel a$ is first order. It occurs at $H_m \sim$2~T at low temperatures and corresponds to a change of magnetic structure from incommensurate to commensurate~\cite{Raymond2007}. This transition is observed up to about 30~T where it merges with $T_N$ giving rise to a tricritical point~\cite{Mishra2021}, above which $T_N$ is still decreasing with field. In Ce$_2$RhIn$_8$, on the contrary, the low-temperature field-induced phase III is very robust. Indeed, although $T_1$ decreases with field above 12~T, it tends to saturate towards 35~T.

When the magnetic field is applied along or close to the $c$ axis in CeRhIn$_5$, a first-order field-induced transition emerges at $H^* \sim$30~T. It was originally observed only for the field tilted away from the $c$ axis and suggested to be of electronic nematic origin~\cite{Ronning2017}. Later on, it was additionally proposed to correspond to a change of magnetic structure from incommensurate to another incommensurate for $H \parallel c$~\cite{Mishra2021} or from commensurate to incommensurate for the field tilted away from the $c$ axis~\cite{Mishra2021a}. Importantly, the phase boundary corresponding to this transition does not cross $T_N(H)$, which continues decreasing above $H^*$~\cite{Mishra2021}. In contrast, in Ce$_2$RhIn$_8$, $T_1$ crosses $T_N$ at about 17~T and increases all the way up to 35~T exceeding the zero field $T_N$.

The field-induced AFM phase III in Ce$_2$RhIn$_8$ is extremely robust against the magnetic field applied both along $a$ and $c$ axis. This is very unusual given that in ordinary antiferromagnets, antiferromagnetic order is continuously suppressed by the magnetic field. Such a behavior is also observed in other Ce-based compounds, although, in some of them, the antiferromagnetic order survives up to considerably higher fields than those of the present study. Furthermore, in these materials, field-induced transitions occur only within the $T_N(H)$ phase boundary. In Ce$_2$RhIn$_8$, on the contrary, $T_1(H)$ crosses $T_N(H)$ for both field orientations. In addition, rather than being suppressed by the magnetic field, $T_1$ increases with field up to 12~T for $H \parallel a$, and all the way up to 35~T for $H \parallel c$. While the exact origin of phase III is unknown, we note that the phase diagram of Ce$_2$RhIn$_8$ for $H \parallel c$ bears a certain resemblance with that of YbCo$_2$Zn$_{20}$~\cite{Honda2014}. In the latter compound, a dome of the field-induced antiferroquadrupole phase is observed between 6 and 21~T below $\sim$0.6~K for $H \parallel$ [111] direction of the cubic structure. Furthermore, an AFM phase develops under pressure above $\simeq$2~GPa. Above this pressure, the field-induced antiferroquadrupole phase persists to temperatures well exceeding $T_N$. While in YbCo$_2$Zn$_{20}$ the antiferroquadrupole phase is observed for $H \parallel$ [111] only, in PrOs$_4$Sb$_{12}$ the field-induced antiferroquadrupole phase is observed for different field directions with a considerable anisotropy~\cite{Tayama2003}. It is therefore an appealing possibility that phase III in Ce$_2$RhIn$_8$ is also of the antiferroquadrupole origin. Additional measurements are required to verify this hypothesis and to elucidate the exact origin of this phase.


In summary, we performed specific-heat measurements in Ce$_2$RhIn$_8$ in static fields up to 35~T applied both along the $a$ and the $c$ axis. For both field directions, a second-order field-induced transition was observed at low temperatures. The corresponding field-induced AFM phase is found to be extremely robust surviving up to 35~T. Microscopic measurements, such as neutron diffraction and NMR, are required to establish the origin of this phase and to comprehend its unusual robustness against magnetic field. Furthermore, additional pulsed field studies are desirable to investigate the behavior of this phase at higher magnetic fields.

\begin{acknowledgments}
We thank M. Zhitomirsky for useful discussions. We acknowledge the support of the LNCMI-CNRS, member of the European Magnetic Field Laboratory (EMFL), and JSPS KAKENHI Grants No. JP22H04933, No. JP20KK0061, and No. JP24H01641.
\end{acknowledgments}

\bibliography{Ce2RhIn8_Phase_Diag}

\begin{thebibliography}{37}%
\makeatletter
\providecommand \@ifxundefined [1]{%
 \@ifx{#1\undefined}
}%
\providecommand \@ifnum [1]{%
 \ifnum #1\expandafter \@firstoftwo
 \else \expandafter \@secondoftwo
 \fi
}%
\providecommand \@ifx [1]{%
 \ifx #1\expandafter \@firstoftwo
 \else \expandafter \@secondoftwo
 \fi
}%
\providecommand \natexlab [1]{#1}%
\providecommand \enquote  [1]{``#1''}%
\providecommand \bibnamefont  [1]{#1}%
\providecommand \bibfnamefont [1]{#1}%
\providecommand \citenamefont [1]{#1}%
\providecommand \href@noop [0]{\@secondoftwo}%
\providecommand \href [0]{\begingroup \@sanitize@url \@href}%
\providecommand \@href[1]{\@@startlink{#1}\@@href}%
\providecommand \@@href[1]{\endgroup#1\@@endlink}%
\providecommand \@sanitize@url [0]{\catcode `\\12\catcode `\$12\catcode
  `\&12\catcode `\#12\catcode `\^12\catcode `\_12\catcode `\%12\relax}%
\providecommand \@@startlink[1]{}%
\providecommand \@@endlink[0]{}%
\providecommand \url  [0]{\begingroup\@sanitize@url \@url }%
\providecommand \@url [1]{\endgroup\@href {#1}{\urlprefix }}%
\providecommand \urlprefix  [0]{URL }%
\providecommand \Eprint [0]{\href }%
\providecommand \doibase [0]{https://doi.org/}%
\providecommand \selectlanguage [0]{\@gobble}%
\providecommand \bibinfo  [0]{\@secondoftwo}%
\providecommand \bibfield  [0]{\@secondoftwo}%
\providecommand \translation [1]{[#1]}%
\providecommand \BibitemOpen [0]{}%
\providecommand \bibitemStop [0]{}%
\providecommand \bibitemNoStop [0]{.\EOS\space}%
\providecommand \EOS [0]{\spacefactor3000\relax}%
\providecommand \BibitemShut  [1]{\csname bibitem#1\endcsname}%
\let\auto@bib@innerbib\@empty
\bibitem [{\citenamefont {Pfleiderer}(2009)}]{Pfleiderer2009}%
  \BibitemOpen
  \bibfield  {author} {\bibinfo {author} {\bibfnamefont {C.}~\bibnamefont
  {Pfleiderer}},\ }\bibfield  {title} {\bibinfo {title} {{Superconducting
  phases of $f$-electron compounds}},\ }\href
  {https://doi.org/10.1103/revmodphys.81.1551} {\bibfield  {journal} {\bibinfo
  {journal} {Rev. Mod. Phys.}\ }\textbf {\bibinfo {volume} {81}},\ \bibinfo
  {pages} {1551} (\bibinfo {year} {2009})}\BibitemShut {NoStop}%
\bibitem [{\citenamefont {Ronning}\ \emph {et~al.}(2017)\citenamefont
  {Ronning}, \citenamefont {Helm}, \citenamefont {Shirer}, \citenamefont
  {Bachmann}, \citenamefont {Balicas}, \citenamefont {Chan}, \citenamefont
  {Ramshaw}, \citenamefont {McDonald}, \citenamefont {Balakirev}, \citenamefont
  {Jaime}, \citenamefont {Bauer},\ and\ \citenamefont {Moll}}]{Ronning2017}%
  \BibitemOpen
  \bibfield  {author} {\bibinfo {author} {\bibfnamefont {F.}~\bibnamefont
  {Ronning}}, \bibinfo {author} {\bibfnamefont {T.}~\bibnamefont {Helm}},
  \bibinfo {author} {\bibfnamefont {K.~R.}\ \bibnamefont {Shirer}}, \bibinfo
  {author} {\bibfnamefont {M.~D.}\ \bibnamefont {Bachmann}}, \bibinfo {author}
  {\bibfnamefont {L.}~\bibnamefont {Balicas}}, \bibinfo {author} {\bibfnamefont
  {M.~K.}\ \bibnamefont {Chan}}, \bibinfo {author} {\bibfnamefont {B.~J.}\
  \bibnamefont {Ramshaw}}, \bibinfo {author} {\bibfnamefont {R.~D.}\
  \bibnamefont {McDonald}}, \bibinfo {author} {\bibfnamefont {F.~F.}\
  \bibnamefont {Balakirev}}, \bibinfo {author} {\bibfnamefont {M.}~\bibnamefont
  {Jaime}}, \bibinfo {author} {\bibfnamefont {E.~D.}\ \bibnamefont {Bauer}},\
  and\ \bibinfo {author} {\bibfnamefont {P.~J.~W.}\ \bibnamefont {Moll}},\
  }\bibfield  {title} {\bibinfo {title} {{Electronic in-plane symmetry breaking
  at field-tuned quantum criticality in CeRhIn$_5$}},\ }\href
  {https://doi.org/10.1038/nature23315} {\bibfield  {journal} {\bibinfo
  {journal} {Nature}\ }\textbf {\bibinfo {volume} {548}},\ \bibinfo {pages}
  {313} (\bibinfo {year} {2017})}\BibitemShut {NoStop}%
\bibitem [{\citenamefont {Mathur}\ \emph {et~al.}(1998)\citenamefont {Mathur},
  \citenamefont {Grosche}, \citenamefont {Julian}, \citenamefont {Walker},
  \citenamefont {Freye}, \citenamefont {Haselwimmer},\ and\ \citenamefont
  {Lonzarich}}]{Mathur1998}%
  \BibitemOpen
  \bibfield  {author} {\bibinfo {author} {\bibfnamefont {N.~D.}\ \bibnamefont
  {Mathur}}, \bibinfo {author} {\bibfnamefont {F.~M.}\ \bibnamefont {Grosche}},
  \bibinfo {author} {\bibfnamefont {S.~R.}\ \bibnamefont {Julian}}, \bibinfo
  {author} {\bibfnamefont {I.~R.}\ \bibnamefont {Walker}}, \bibinfo {author}
  {\bibfnamefont {D.~M.}\ \bibnamefont {Freye}}, \bibinfo {author}
  {\bibfnamefont {R.~K.~W.}\ \bibnamefont {Haselwimmer}},\ and\ \bibinfo
  {author} {\bibfnamefont {G.~G.}\ \bibnamefont {Lonzarich}},\ }\bibfield
  {title} {\bibinfo {title} {Magnetically mediated superconductivity in heavy
  fermion compounds},\ }\href {https://doi.org/10.1038/27838} {\bibfield
  {journal} {\bibinfo  {journal} {Nature}\ }\textbf {\bibinfo {volume} {394}},\
  \bibinfo {pages} {39} (\bibinfo {year} {1998})}\BibitemShut {NoStop}%
\bibitem [{\citenamefont {Hegger}\ \emph {et~al.}(2000)\citenamefont {Hegger},
  \citenamefont {Petrovic}, \citenamefont {Moshopoulou}, \citenamefont
  {Hundley}, \citenamefont {Sarrao}, \citenamefont {Fisk},\ and\ \citenamefont
  {Thompson}}]{Hegger2000}%
  \BibitemOpen
  \bibfield  {author} {\bibinfo {author} {\bibfnamefont {H.}~\bibnamefont
  {Hegger}}, \bibinfo {author} {\bibfnamefont {C.}~\bibnamefont {Petrovic}},
  \bibinfo {author} {\bibfnamefont {E.~G.}\ \bibnamefont {Moshopoulou}},
  \bibinfo {author} {\bibfnamefont {M.~F.}\ \bibnamefont {Hundley}}, \bibinfo
  {author} {\bibfnamefont {J.~L.}\ \bibnamefont {Sarrao}}, \bibinfo {author}
  {\bibfnamefont {Z.}~\bibnamefont {Fisk}},\ and\ \bibinfo {author}
  {\bibfnamefont {J.~D.}\ \bibnamefont {Thompson}},\ }\bibfield  {title}
  {\bibinfo {title} {{Pressure-Induced Superconductivity in Quasi-2D
  CeRhIn$_5$}},\ }\href {https://doi.org/10.1103/physrevlett.84.4986}
  {\bibfield  {journal} {\bibinfo  {journal} {Phys. Rev. Lett.}\ }\textbf
  {\bibinfo {volume} {84}},\ \bibinfo {pages} {4986} (\bibinfo {year}
  {2000})}\BibitemShut {NoStop}%
\bibitem [{\citenamefont {Bauer}\ \emph {et~al.}(2010)\citenamefont {Bauer},
  \citenamefont {Lee}, \citenamefont {Sidorov}, \citenamefont {Kurita},
  \citenamefont {Gofryk}, \citenamefont {Zhu}, \citenamefont {Ronning},
  \citenamefont {Movshovich}, \citenamefont {Thompson},\ and\ \citenamefont
  {Park}}]{Bauer2010a}%
  \BibitemOpen
  \bibfield  {author} {\bibinfo {author} {\bibfnamefont {E.~D.}\ \bibnamefont
  {Bauer}}, \bibinfo {author} {\bibfnamefont {H.~O.}\ \bibnamefont {Lee}},
  \bibinfo {author} {\bibfnamefont {V.~A.}\ \bibnamefont {Sidorov}}, \bibinfo
  {author} {\bibfnamefont {N.}~\bibnamefont {Kurita}}, \bibinfo {author}
  {\bibfnamefont {K.}~\bibnamefont {Gofryk}}, \bibinfo {author} {\bibfnamefont
  {J.-X.}\ \bibnamefont {Zhu}}, \bibinfo {author} {\bibfnamefont
  {F.}~\bibnamefont {Ronning}}, \bibinfo {author} {\bibfnamefont
  {R.}~\bibnamefont {Movshovich}}, \bibinfo {author} {\bibfnamefont {J.~D.}\
  \bibnamefont {Thompson}},\ and\ \bibinfo {author} {\bibfnamefont
  {T.}~\bibnamefont {Park}},\ }\bibfield  {title} {\bibinfo {title}
  {{Pressure-induced superconducting state and effective mass enhancement near
  the antiferromagnetic quantum critical point of CePt$_2$In$_7$}},\ }\href
  {https://doi.org/10.1103/physrevb.81.180507} {\bibfield  {journal} {\bibinfo
  {journal} {Phys. Rev. B}\ }\textbf {\bibinfo {volume} {81}},\ \bibinfo
  {pages} {180507} (\bibinfo {year} {2010})}\BibitemShut {NoStop}%
\bibitem [{\citenamefont {Ebihara}\ \emph {et~al.}(2004)\citenamefont
  {Ebihara}, \citenamefont {Harrison}, \citenamefont {Jaime}, \citenamefont
  {Uji},\ and\ \citenamefont {Lashley}}]{Ebihara2004}%
  \BibitemOpen
  \bibfield  {author} {\bibinfo {author} {\bibfnamefont {T.}~\bibnamefont
  {Ebihara}}, \bibinfo {author} {\bibfnamefont {N.}~\bibnamefont {Harrison}},
  \bibinfo {author} {\bibfnamefont {M.}~\bibnamefont {Jaime}}, \bibinfo
  {author} {\bibfnamefont {S.}~\bibnamefont {Uji}},\ and\ \bibinfo {author}
  {\bibfnamefont {J.~C.}\ \bibnamefont {Lashley}},\ }\bibfield  {title}
  {\bibinfo {title} {{Emergent Fluctuation Hot Spots on the Fermi Surface of
  CeIn$_3$ in Strong Magnetic Fields}},\ }\href
  {https://doi.org/10.1103/physrevlett.93.246401} {\bibfield  {journal}
  {\bibinfo  {journal} {Phys. Rev. Lett.}\ }\textbf {\bibinfo {volume} {93}},\
  \bibinfo {pages} {246401} (\bibinfo {year} {2004})}\BibitemShut {NoStop}%
\bibitem [{\citenamefont {Moll}\ \emph {et~al.}(2017)\citenamefont {Moll},
  \citenamefont {Helm}, \citenamefont {Zhang}, \citenamefont {Batista},
  \citenamefont {Harrison}, \citenamefont {McDonald}, \citenamefont {Winter},
  \citenamefont {Ramshaw}, \citenamefont {Chan}, \citenamefont {Balakirev},
  \citenamefont {Batlogg}, \citenamefont {Bauer},\ and\ \citenamefont
  {Ronning}}]{Moll2017}%
  \BibitemOpen
  \bibfield  {author} {\bibinfo {author} {\bibfnamefont {P.~J.~W.}\
  \bibnamefont {Moll}}, \bibinfo {author} {\bibfnamefont {T.}~\bibnamefont
  {Helm}}, \bibinfo {author} {\bibfnamefont {S.-S.}\ \bibnamefont {Zhang}},
  \bibinfo {author} {\bibfnamefont {C.~D.}\ \bibnamefont {Batista}}, \bibinfo
  {author} {\bibfnamefont {N.}~\bibnamefont {Harrison}}, \bibinfo {author}
  {\bibfnamefont {R.~D.}\ \bibnamefont {McDonald}}, \bibinfo {author}
  {\bibfnamefont {L.~E.}\ \bibnamefont {Winter}}, \bibinfo {author}
  {\bibfnamefont {B.~J.}\ \bibnamefont {Ramshaw}}, \bibinfo {author}
  {\bibfnamefont {M.~K.}\ \bibnamefont {Chan}}, \bibinfo {author}
  {\bibfnamefont {F.~F.}\ \bibnamefont {Balakirev}}, \bibinfo {author}
  {\bibfnamefont {B.}~\bibnamefont {Batlogg}}, \bibinfo {author} {\bibfnamefont
  {E.~D.}\ \bibnamefont {Bauer}},\ and\ \bibinfo {author} {\bibfnamefont
  {F.}~\bibnamefont {Ronning}},\ }\bibfield  {title} {\bibinfo {title}
  {{Emergent magnetic anisotropy in the cubic heavy-fermion metal CeIn$_3$}},\
  }\href {https://doi.org/10.1038/s41535-017-0052-5} {\bibfield  {journal}
  {\bibinfo  {journal} {npj Quantum Materials}\ }\textbf {\bibinfo {volume}
  {2}},\ \bibinfo {pages} {46} (\bibinfo {year} {2017})}\BibitemShut {NoStop}%
\bibitem [{\citenamefont {Jiao}\ \emph {et~al.}(2015)\citenamefont {Jiao},
  \citenamefont {Chen}, \citenamefont {Kohama}, \citenamefont {Graf},
  \citenamefont {Bauer}, \citenamefont {Singleton}, \citenamefont {Zhu},
  \citenamefont {Weng}, \citenamefont {Pang}, \citenamefont {Shang},
  \citenamefont {Zhang}, \citenamefont {Lee}, \citenamefont {Park},
  \citenamefont {Jaime}, \citenamefont {Thompson}, \citenamefont {Steglich},
  \citenamefont {Si},\ and\ \citenamefont {Yuan}}]{Jiao2015}%
  \BibitemOpen
  \bibfield  {author} {\bibinfo {author} {\bibfnamefont {L.}~\bibnamefont
  {Jiao}}, \bibinfo {author} {\bibfnamefont {Y.}~\bibnamefont {Chen}}, \bibinfo
  {author} {\bibfnamefont {Y.}~\bibnamefont {Kohama}}, \bibinfo {author}
  {\bibfnamefont {D.}~\bibnamefont {Graf}}, \bibinfo {author} {\bibfnamefont
  {E.~D.}\ \bibnamefont {Bauer}}, \bibinfo {author} {\bibfnamefont
  {J.}~\bibnamefont {Singleton}}, \bibinfo {author} {\bibfnamefont {J.-X.}\
  \bibnamefont {Zhu}}, \bibinfo {author} {\bibfnamefont {Z.}~\bibnamefont
  {Weng}}, \bibinfo {author} {\bibfnamefont {G.}~\bibnamefont {Pang}}, \bibinfo
  {author} {\bibfnamefont {T.}~\bibnamefont {Shang}}, \bibinfo {author}
  {\bibfnamefont {J.}~\bibnamefont {Zhang}}, \bibinfo {author} {\bibfnamefont
  {H.-O.}\ \bibnamefont {Lee}}, \bibinfo {author} {\bibfnamefont
  {T.}~\bibnamefont {Park}}, \bibinfo {author} {\bibfnamefont {M.}~\bibnamefont
  {Jaime}}, \bibinfo {author} {\bibfnamefont {J.~D.}\ \bibnamefont {Thompson}},
  \bibinfo {author} {\bibfnamefont {F.}~\bibnamefont {Steglich}}, \bibinfo
  {author} {\bibfnamefont {Q.}~\bibnamefont {Si}},\ and\ \bibinfo {author}
  {\bibfnamefont {H.~Q.}\ \bibnamefont {Yuan}},\ }\bibfield  {title} {\bibinfo
  {title} {{Fermi surface reconstruction and multiple quantum phase transitions
  in the antiferromagnet CeRhIn$_5$}},\ }\href
  {https://doi.org/10.1073/pnas.1413932112} {\bibfield  {journal} {\bibinfo
  {journal} {Proc. Natl. Acad. Sci. USA}\ }\textbf {\bibinfo {volume} {112}},\
  \bibinfo {pages} {673} (\bibinfo {year} {2015})}\BibitemShut {NoStop}%
\bibitem [{\citenamefont {Jiao}\ \emph {et~al.}(2019)\citenamefont {Jiao},
  \citenamefont {Smidman}, \citenamefont {Kohama}, \citenamefont {Wang},
  \citenamefont {Graf}, \citenamefont {Weng}, \citenamefont {Zhang},
  \citenamefont {Matsuo}, \citenamefont {Bauer}, \citenamefont {Lee},
  \citenamefont {Kirchner}, \citenamefont {Singleton}, \citenamefont {Kindo},
  \citenamefont {Wosnitza}, \citenamefont {Steglich}, \citenamefont
  {Thompson},\ and\ \citenamefont {Yuan}}]{Jiao2019}%
  \BibitemOpen
  \bibfield  {author} {\bibinfo {author} {\bibfnamefont {L.}~\bibnamefont
  {Jiao}}, \bibinfo {author} {\bibfnamefont {M.}~\bibnamefont {Smidman}},
  \bibinfo {author} {\bibfnamefont {Y.}~\bibnamefont {Kohama}}, \bibinfo
  {author} {\bibfnamefont {Z.~S.}\ \bibnamefont {Wang}}, \bibinfo {author}
  {\bibfnamefont {D.}~\bibnamefont {Graf}}, \bibinfo {author} {\bibfnamefont
  {Z.~F.}\ \bibnamefont {Weng}}, \bibinfo {author} {\bibfnamefont {Y.~J.}\
  \bibnamefont {Zhang}}, \bibinfo {author} {\bibfnamefont {A.}~\bibnamefont
  {Matsuo}}, \bibinfo {author} {\bibfnamefont {E.~D.}\ \bibnamefont {Bauer}},
  \bibinfo {author} {\bibfnamefont {H.}~\bibnamefont {Lee}}, \bibinfo {author}
  {\bibfnamefont {S.}~\bibnamefont {Kirchner}}, \bibinfo {author}
  {\bibfnamefont {J.}~\bibnamefont {Singleton}}, \bibinfo {author}
  {\bibfnamefont {K.}~\bibnamefont {Kindo}}, \bibinfo {author} {\bibfnamefont
  {J.}~\bibnamefont {Wosnitza}}, \bibinfo {author} {\bibfnamefont
  {F.}~\bibnamefont {Steglich}}, \bibinfo {author} {\bibfnamefont {J.~D.}\
  \bibnamefont {Thompson}},\ and\ \bibinfo {author} {\bibfnamefont {H.~Q.}\
  \bibnamefont {Yuan}},\ }\bibfield  {title} {\bibinfo {title} {{Enhancement of
  the effective mass at high magnetic fields in CeRhIn$_5$}},\ }\href
  {https://doi.org/10.1103/physrevb.99.045127} {\bibfield  {journal} {\bibinfo
  {journal} {Phys. Rev. B}\ }\textbf {\bibinfo {volume} {99}},\ \bibinfo
  {pages} {045127} (\bibinfo {year} {2019})}\BibitemShut {NoStop}%
\bibitem [{\citenamefont {Krupko}\ \emph {et~al.}(2016)\citenamefont {Krupko},
  \citenamefont {Demuer}, \citenamefont {Ota}, \citenamefont {Hirose},
  \citenamefont {Settai},\ and\ \citenamefont {Sheikin}}]{Krupko2016}%
  \BibitemOpen
  \bibfield  {author} {\bibinfo {author} {\bibfnamefont {Y.}~\bibnamefont
  {Krupko}}, \bibinfo {author} {\bibfnamefont {A.}~\bibnamefont {Demuer}},
  \bibinfo {author} {\bibfnamefont {S.}~\bibnamefont {Ota}}, \bibinfo {author}
  {\bibfnamefont {Y.}~\bibnamefont {Hirose}}, \bibinfo {author} {\bibfnamefont
  {R.}~\bibnamefont {Settai}},\ and\ \bibinfo {author} {\bibfnamefont
  {I.}~\bibnamefont {Sheikin}},\ }\bibfield  {title} {\bibinfo {title}
  {{Specific heat in high magnetic fields and magnetic phase diagram of
  CePt$_2$In$_7$}},\ }\href {https://doi.org/10.1103/physrevb.93.085121}
  {\bibfield  {journal} {\bibinfo  {journal} {Phys. Rev. B}\ }\textbf {\bibinfo
  {volume} {93}},\ \bibinfo {pages} {085121} (\bibinfo {year}
  {2016})}\BibitemShut {NoStop}%
\bibitem [{\citenamefont {Mishra}\ \emph
  {et~al.}(2021{\natexlab{a}})\citenamefont {Mishra}, \citenamefont {Demuer},
  \citenamefont {Aoki},\ and\ \citenamefont {Sheikin}}]{Mishra2021}%
  \BibitemOpen
  \bibfield  {author} {\bibinfo {author} {\bibfnamefont {S.}~\bibnamefont
  {Mishra}}, \bibinfo {author} {\bibfnamefont {A.}~\bibnamefont {Demuer}},
  \bibinfo {author} {\bibfnamefont {D.}~\bibnamefont {Aoki}},\ and\ \bibinfo
  {author} {\bibfnamefont {I.}~\bibnamefont {Sheikin}},\ }\bibfield  {title}
  {\bibinfo {title} {{Specific heat of CeRhIn$_5$ in high magnetic fields:
  Magnetic phase diagram revisited}},\ }\href
  {https://doi.org/10.1103/physrevb.103.045110} {\bibfield  {journal} {\bibinfo
   {journal} {Phys. Rev. B}\ }\textbf {\bibinfo {volume} {103}},\ \bibinfo
  {pages} {045110} (\bibinfo {year} {2021}{\natexlab{a}})}\BibitemShut
  {NoStop}%
\bibitem [{\citenamefont {Moll}\ \emph {et~al.}(2015)\citenamefont {Moll},
  \citenamefont {Zeng}, \citenamefont {Balicas}, \citenamefont {Galeski},
  \citenamefont {Balakirev}, \citenamefont {Bauer},\ and\ \citenamefont
  {Ronning}}]{Moll2015}%
  \BibitemOpen
  \bibfield  {author} {\bibinfo {author} {\bibfnamefont {P.~J.~W.}\
  \bibnamefont {Moll}}, \bibinfo {author} {\bibfnamefont {B.}~\bibnamefont
  {Zeng}}, \bibinfo {author} {\bibfnamefont {L.}~\bibnamefont {Balicas}},
  \bibinfo {author} {\bibfnamefont {S.}~\bibnamefont {Galeski}}, \bibinfo
  {author} {\bibfnamefont {F.~F.}\ \bibnamefont {Balakirev}}, \bibinfo {author}
  {\bibfnamefont {E.~D.}\ \bibnamefont {Bauer}},\ and\ \bibinfo {author}
  {\bibfnamefont {F.}~\bibnamefont {Ronning}},\ }\bibfield  {title} {\bibinfo
  {title} {{Field-induced density wave in the heavy-fermion compound
  CeRhIn$_5$}},\ }\href {https://doi.org/10.1038/ncomms7663} {\bibfield
  {journal} {\bibinfo  {journal} {Nat. Commun.}\ }\textbf {\bibinfo {volume}
  {6}},\ \bibinfo {pages} {6663} (\bibinfo {year} {2015})}\BibitemShut
  {NoStop}%
\bibitem [{\citenamefont {Rosa}\ \emph {et~al.}(2019)\citenamefont {Rosa},
  \citenamefont {Thomas}, \citenamefont {Balakirev}, \citenamefont {Bauer},
  \citenamefont {Fernandes}, \citenamefont {Thompson}, \citenamefont
  {Ronning},\ and\ \citenamefont {Jaime}}]{Rosa2019}%
  \BibitemOpen
  \bibfield  {author} {\bibinfo {author} {\bibfnamefont {P.}~\bibnamefont
  {Rosa}}, \bibinfo {author} {\bibfnamefont {S.}~\bibnamefont {Thomas}},
  \bibinfo {author} {\bibfnamefont {F.}~\bibnamefont {Balakirev}}, \bibinfo
  {author} {\bibfnamefont {E.}~\bibnamefont {Bauer}}, \bibinfo {author}
  {\bibfnamefont {R.}~\bibnamefont {Fernandes}}, \bibinfo {author}
  {\bibfnamefont {J.}~\bibnamefont {Thompson}}, \bibinfo {author}
  {\bibfnamefont {F.}~\bibnamefont {Ronning}},\ and\ \bibinfo {author}
  {\bibfnamefont {M.}~\bibnamefont {Jaime}},\ }\bibfield  {title} {\bibinfo
  {title} {{Enhanced Hybridization Sets the Stage for Electronic Nematicity in
  CeRhIn$_5$}},\ }\href {https://doi.org/10.1103/physrevlett.122.016402}
  {\bibfield  {journal} {\bibinfo  {journal} {Phys. Rev. Lett.}\ }\textbf
  {\bibinfo {volume} {122}},\ \bibinfo {pages} {016402} (\bibinfo {year}
  {2019})}\BibitemShut {NoStop}%
\bibitem [{\citenamefont {Kurihara}\ \emph {et~al.}(2020)\citenamefont
  {Kurihara}, \citenamefont {Miyake}, \citenamefont {Tokunaga}, \citenamefont
  {Hirose},\ and\ \citenamefont {Settai}}]{Kurihara2020}%
  \BibitemOpen
  \bibfield  {author} {\bibinfo {author} {\bibfnamefont {R.}~\bibnamefont
  {Kurihara}}, \bibinfo {author} {\bibfnamefont {A.}~\bibnamefont {Miyake}},
  \bibinfo {author} {\bibfnamefont {M.}~\bibnamefont {Tokunaga}}, \bibinfo
  {author} {\bibfnamefont {Y.}~\bibnamefont {Hirose}},\ and\ \bibinfo {author}
  {\bibfnamefont {R.}~\bibnamefont {Settai}},\ }\bibfield  {title} {\bibinfo
  {title} {{High-field ultrasonic study of quadrupole ordering and crystal
  symmetry breaking in CeRhIn$_5$}},\ }\href
  {https://doi.org/10.1103/physrevb.101.155125} {\bibfield  {journal} {\bibinfo
   {journal} {Phys. Rev. B}\ }\textbf {\bibinfo {volume} {101}},\ \bibinfo
  {pages} {155125} (\bibinfo {year} {2020})}\BibitemShut {NoStop}%
\bibitem [{\citenamefont {Lesseux}\ \emph {et~al.}(2020)\citenamefont
  {Lesseux}, \citenamefont {Sakai}, \citenamefont {Hattori}, \citenamefont
  {Tokunaga}, \citenamefont {Kambe}, \citenamefont {Kuhns}, \citenamefont
  {Reyes}, \citenamefont {Thompson}, \citenamefont {Pagliuso},\ and\
  \citenamefont {Urbano}}]{Lesseux2020}%
  \BibitemOpen
  \bibfield  {author} {\bibinfo {author} {\bibfnamefont {G.~G.}\ \bibnamefont
  {Lesseux}}, \bibinfo {author} {\bibfnamefont {H.}~\bibnamefont {Sakai}},
  \bibinfo {author} {\bibfnamefont {T.}~\bibnamefont {Hattori}}, \bibinfo
  {author} {\bibfnamefont {Y.}~\bibnamefont {Tokunaga}}, \bibinfo {author}
  {\bibfnamefont {S.}~\bibnamefont {Kambe}}, \bibinfo {author} {\bibfnamefont
  {P.~L.}\ \bibnamefont {Kuhns}}, \bibinfo {author} {\bibfnamefont {A.~P.}\
  \bibnamefont {Reyes}}, \bibinfo {author} {\bibfnamefont {J.~D.}\ \bibnamefont
  {Thompson}}, \bibinfo {author} {\bibfnamefont {P.~G.}\ \bibnamefont
  {Pagliuso}},\ and\ \bibinfo {author} {\bibfnamefont {R.~R.}\ \bibnamefont
  {Urbano}},\ }\bibfield  {title} {\bibinfo {title} {{Orbitally defined
  field-induced electronic state in a Kondo lattice}},\ }\href
  {https://doi.org/10.1103/physrevb.101.165111} {\bibfield  {journal} {\bibinfo
   {journal} {Phys. Rev. B}\ }\textbf {\bibinfo {volume} {101}},\ \bibinfo
  {pages} {165111} (\bibinfo {year} {2020})}\BibitemShut {NoStop}%
\bibitem [{\citenamefont {Mishra}\ \emph
  {et~al.}(2021{\natexlab{b}})\citenamefont {Mishra}, \citenamefont {Gorbunov},
  \citenamefont {Campbell}, \citenamefont {LeBoeuf}, \citenamefont {Hornung},
  \citenamefont {Klotz}, \citenamefont {Zherlitsyn}, \citenamefont {Harima},
  \citenamefont {Wosnitza}, \citenamefont {Aoki}, \citenamefont {McCollam},\
  and\ \citenamefont {Sheikin}}]{Mishra2021a}%
  \BibitemOpen
  \bibfield  {author} {\bibinfo {author} {\bibfnamefont {S.}~\bibnamefont
  {Mishra}}, \bibinfo {author} {\bibfnamefont {D.}~\bibnamefont {Gorbunov}},
  \bibinfo {author} {\bibfnamefont {D.~J.}\ \bibnamefont {Campbell}}, \bibinfo
  {author} {\bibfnamefont {D.}~\bibnamefont {LeBoeuf}}, \bibinfo {author}
  {\bibfnamefont {J.}~\bibnamefont {Hornung}}, \bibinfo {author} {\bibfnamefont
  {J.}~\bibnamefont {Klotz}}, \bibinfo {author} {\bibfnamefont
  {S.}~\bibnamefont {Zherlitsyn}}, \bibinfo {author} {\bibfnamefont
  {H.}~\bibnamefont {Harima}}, \bibinfo {author} {\bibfnamefont
  {J.}~\bibnamefont {Wosnitza}}, \bibinfo {author} {\bibfnamefont
  {D.}~\bibnamefont {Aoki}}, \bibinfo {author} {\bibfnamefont {A.}~\bibnamefont
  {McCollam}},\ and\ \bibinfo {author} {\bibfnamefont {I.}~\bibnamefont
  {Sheikin}},\ }\bibfield  {title} {\bibinfo {title} {{Origin of the 30 T
  transition in CeRhIn$_5$ in tilted magnetic fields}},\ }\href
  {https://doi.org/10.1103/physrevb.103.165124} {\bibfield  {journal} {\bibinfo
   {journal} {Phys. Rev. B}\ }\textbf {\bibinfo {volume} {103}},\ \bibinfo
  {pages} {165124} (\bibinfo {year} {2021}{\natexlab{b}})}\BibitemShut
  {NoStop}%
\bibitem [{\citenamefont {Cornelius}\ \emph {et~al.}(2001)\citenamefont
  {Cornelius}, \citenamefont {Pagliuso}, \citenamefont {Hundley},\ and\
  \citenamefont {Sarrao}}]{Cornelius2001}%
  \BibitemOpen
  \bibfield  {author} {\bibinfo {author} {\bibfnamefont {A.~L.}\ \bibnamefont
  {Cornelius}}, \bibinfo {author} {\bibfnamefont {P.~G.}\ \bibnamefont
  {Pagliuso}}, \bibinfo {author} {\bibfnamefont {M.~F.}\ \bibnamefont
  {Hundley}},\ and\ \bibinfo {author} {\bibfnamefont {J.~L.}\ \bibnamefont
  {Sarrao}},\ }\bibfield  {title} {\bibinfo {title} {{Field-induced magnetic
  transitions in the quasi-two-dimensional heavy-fermion antiferromagnets
  Ce$_n$RhIn$_{3n+2}$ ($n =$ 1 or 2)}},\ }\href
  {https://doi.org/10.1103/physrevb.64.144411} {\bibfield  {journal} {\bibinfo
  {journal} {Phys. Rev. B}\ }\textbf {\bibinfo {volume} {64}},\ \bibinfo
  {pages} {144411} (\bibinfo {year} {2001})}\BibitemShut {NoStop}%
\bibitem [{\citenamefont {Kratochvílová}\ \emph {et~al.}(2015)\citenamefont
  {Kratochvílová}, \citenamefont {Prokleška}, \citenamefont {Uhlířová},
  \citenamefont {Tkáč}, \citenamefont {Dušek}, \citenamefont {Sechovský},\
  and\ \citenamefont {Custers}}]{Kratochvilova2015a}%
  \BibitemOpen
  \bibfield  {author} {\bibinfo {author} {\bibfnamefont {M.}~\bibnamefont
  {Kratochvílová}}, \bibinfo {author} {\bibfnamefont {J.}~\bibnamefont
  {Prokleška}}, \bibinfo {author} {\bibfnamefont {K.}~\bibnamefont
  {Uhlířová}}, \bibinfo {author} {\bibfnamefont {V.}~\bibnamefont {Tkáč}},
  \bibinfo {author} {\bibfnamefont {M.}~\bibnamefont {Dušek}}, \bibinfo
  {author} {\bibfnamefont {V.}~\bibnamefont {Sechovský}},\ and\ \bibinfo
  {author} {\bibfnamefont {J.}~\bibnamefont {Custers}},\ }\bibfield  {title}
  {\bibinfo {title} {{Coexistence of Antiferromagnetism and Superconductivity
  in Heavy Fermion Cerium Compound Ce$_3$PdIn$_{11}$}},\ }\href
  {https://doi.org/10.1038/srep15904} {\bibfield  {journal} {\bibinfo
  {journal} {Sci. Rep.}\ }\textbf {\bibinfo {volume} {5}},\ \bibinfo {pages}
  {15904} (\bibinfo {year} {2015})}\BibitemShut {NoStop}%
\bibitem [{\citenamefont {Prokleška}\ \emph {et~al.}(2015)\citenamefont
  {Prokleška}, \citenamefont {Kratochvílová}, \citenamefont {Uhlířová},
  \citenamefont {Sechovský},\ and\ \citenamefont {Custers}}]{Prokleska2015}%
  \BibitemOpen
  \bibfield  {author} {\bibinfo {author} {\bibfnamefont {J.}~\bibnamefont
  {Prokleška}}, \bibinfo {author} {\bibfnamefont {M.}~\bibnamefont
  {Kratochvílová}}, \bibinfo {author} {\bibfnamefont {K.}~\bibnamefont
  {Uhlířová}}, \bibinfo {author} {\bibfnamefont {V.}~\bibnamefont
  {Sechovský}},\ and\ \bibinfo {author} {\bibfnamefont {J.}~\bibnamefont
  {Custers}},\ }\bibfield  {title} {\bibinfo {title} {{Magnetism,
  superconductivity, and quantum criticality in the multisite cerium
  heavy-fermion compound Ce$_3$PtIn$_{11}$}},\ }\href
  {https://doi.org/10.1103/physrevb.92.161114} {\bibfield  {journal} {\bibinfo
  {journal} {Phys. Rev. B}\ }\textbf {\bibinfo {volume} {92}},\ \bibinfo
  {pages} {161114} (\bibinfo {year} {2015})}\BibitemShut {NoStop}%
\bibitem [{\citenamefont {Das}\ \emph {et~al.}(2019)\citenamefont {Das},
  \citenamefont {Gnida},\ and\ \citenamefont {Kaczorowski}}]{Das2019}%
  \BibitemOpen
  \bibfield  {author} {\bibinfo {author} {\bibfnamefont {D.}~\bibnamefont
  {Das}}, \bibinfo {author} {\bibfnamefont {D.}~\bibnamefont {Gnida}},\ and\
  \bibinfo {author} {\bibfnamefont {D.}~\bibnamefont {Kaczorowski}},\
  }\bibfield  {title} {\bibinfo {title} {{Anisotropic magnetotransport and
  magnetic phase diagrams of the antiferromagnetic heavy-fermion superconductor
  Ce$_3$PdIn$_{11}$}},\ }\href {https://doi.org/10.1103/physrevb.99.054425}
  {\bibfield  {journal} {\bibinfo  {journal} {Phys. Rev. B}\ }\textbf {\bibinfo
  {volume} {99}},\ \bibinfo {pages} {054425} (\bibinfo {year}
  {2019})}\BibitemShut {NoStop}%
\bibitem [{\citenamefont {Das}\ \emph {et~al.}(2018)\citenamefont {Das},
  \citenamefont {Gnida}, \citenamefont {Bochenek}, \citenamefont {Rudenko},
  \citenamefont {Daszkiewicz},\ and\ \citenamefont {Kaczorowski}}]{Das2018}%
  \BibitemOpen
  \bibfield  {author} {\bibinfo {author} {\bibfnamefont {D.}~\bibnamefont
  {Das}}, \bibinfo {author} {\bibfnamefont {D.}~\bibnamefont {Gnida}}, \bibinfo
  {author} {\bibfnamefont {{\L}.}~\bibnamefont {Bochenek}}, \bibinfo {author}
  {\bibfnamefont {A.}~\bibnamefont {Rudenko}}, \bibinfo {author} {\bibfnamefont
  {M.}~\bibnamefont {Daszkiewicz}},\ and\ \bibinfo {author} {\bibfnamefont
  {D.}~\bibnamefont {Kaczorowski}},\ }\bibfield  {title} {\bibinfo {title}
  {{Magnetic field driven complex phase diagram of antiferromagnetic
  heavy-fermion superconductor Ce$_3$PtIn$_{11}$}},\ }\href
  {https://doi.org/10.1038/s41598-018-34991-7} {\bibfield  {journal} {\bibinfo
  {journal} {Sci. Rep.}\ }\textbf {\bibinfo {volume} {8}},\ \bibinfo {pages}
  {16703} (\bibinfo {year} {2018})}\BibitemShut {NoStop}%
\bibitem [{\citenamefont {Grin’}\ \emph {et~al.}(1979)\citenamefont
  {Grin’}, \citenamefont {Yarmolyuk},\ and\ \citenamefont
  {Gladyshevskii}}]{Grin’1979}%
  \BibitemOpen
  \bibfield  {author} {\bibinfo {author} {\bibfnamefont {Y.}~\bibnamefont
  {Grin’}}, \bibinfo {author} {\bibfnamefont {Y.}~\bibnamefont {Yarmolyuk}},\
  and\ \bibinfo {author} {\bibfnamefont {E.}~\bibnamefont {Gladyshevskii}},\
  }\bibfield  {title} {\bibinfo {title} {{Crystal structures of $R_2$CoGa$_8$
  compounds ($R$ = Sm, Gd, Tb, Dy, Ho, Er, Tm, Lu, Y) and $R$CoGa$_5$ compounds
  ($R$ = Gd, Tb, Dy, Ho, Er, Tm, Lu, or Y)}},\ }\href@noop {} {\bibfield
  {journal} {\bibinfo  {journal} {Sov. Phys. Crystallogr.}\ }\textbf {\bibinfo
  {volume} {24}},\ \bibinfo {pages} {137} (\bibinfo {year} {1979})}\BibitemShut
  {NoStop}%
\bibitem [{\citenamefont {Thompson}\ \emph {et~al.}(2001)\citenamefont
  {Thompson}, \citenamefont {Movshovich}, \citenamefont {Fisk}, \citenamefont
  {Bouquet}, \citenamefont {Curro}, \citenamefont {Fisher}, \citenamefont
  {Hammel}, \citenamefont {Hegger}, \citenamefont {Hundley}, \citenamefont
  {Jaime}, \citenamefont {Pagliuso}, \citenamefont {Petrovic}, \citenamefont
  {Phillips},\ and\ \citenamefont {Sarrao}}]{Thompson2001}%
  \BibitemOpen
  \bibfield  {author} {\bibinfo {author} {\bibfnamefont {J.}~\bibnamefont
  {Thompson}}, \bibinfo {author} {\bibfnamefont {R.}~\bibnamefont
  {Movshovich}}, \bibinfo {author} {\bibfnamefont {Z.}~\bibnamefont {Fisk}},
  \bibinfo {author} {\bibfnamefont {F.}~\bibnamefont {Bouquet}}, \bibinfo
  {author} {\bibfnamefont {N.}~\bibnamefont {Curro}}, \bibinfo {author}
  {\bibfnamefont {R.}~\bibnamefont {Fisher}}, \bibinfo {author} {\bibfnamefont
  {P.}~\bibnamefont {Hammel}}, \bibinfo {author} {\bibfnamefont
  {H.}~\bibnamefont {Hegger}}, \bibinfo {author} {\bibfnamefont
  {M.}~\bibnamefont {Hundley}}, \bibinfo {author} {\bibfnamefont
  {M.}~\bibnamefont {Jaime}}, \bibinfo {author} {\bibfnamefont
  {P.}~\bibnamefont {Pagliuso}}, \bibinfo {author} {\bibfnamefont
  {C.}~\bibnamefont {Petrovic}}, \bibinfo {author} {\bibfnamefont
  {N.}~\bibnamefont {Phillips}},\ and\ \bibinfo {author} {\bibfnamefont
  {J.}~\bibnamefont {Sarrao}},\ }\bibfield  {title} {\bibinfo {title}
  {Superconductivity and magnetism in a new class of heavy-fermion materials},\
  }\href {https://doi.org/10.1016/s0304-8853(00)00602-8} {\bibfield  {journal}
  {\bibinfo  {journal} {J. Magn. Magn. Mater.}\ }\textbf {\bibinfo {volume}
  {226–230}},\ \bibinfo {pages} {5} (\bibinfo {year} {2001})}\BibitemShut
  {NoStop}%
\bibitem [{\citenamefont {Bao}\ \emph {et~al.}(2001)\citenamefont {Bao},
  \citenamefont {Pagliuso}, \citenamefont {Sarrao}, \citenamefont {Thompson},
  \citenamefont {Fisk},\ and\ \citenamefont {Lynn}}]{Bao2001}%
  \BibitemOpen
  \bibfield  {author} {\bibinfo {author} {\bibfnamefont {W.}~\bibnamefont
  {Bao}}, \bibinfo {author} {\bibfnamefont {P.~G.}\ \bibnamefont {Pagliuso}},
  \bibinfo {author} {\bibfnamefont {J.~L.}\ \bibnamefont {Sarrao}}, \bibinfo
  {author} {\bibfnamefont {J.~D.}\ \bibnamefont {Thompson}}, \bibinfo {author}
  {\bibfnamefont {Z.}~\bibnamefont {Fisk}},\ and\ \bibinfo {author}
  {\bibfnamefont {J.~W.}\ \bibnamefont {Lynn}},\ }\bibfield  {title} {\bibinfo
  {title} {{Magnetic structure of heavy-fermion Ce$_2$RhIn$_8$}},\ }\href
  {https://doi.org/10.1103/physrevb.64.020401} {\bibfield  {journal} {\bibinfo
  {journal} {Phys. Rev. B}\ }\textbf {\bibinfo {volume} {64}},\ \bibinfo
  {pages} {020401} (\bibinfo {year} {2001})}\BibitemShut {NoStop}%
\bibitem [{\citenamefont {Ueda}\ \emph {et~al.}(2004)\citenamefont {Ueda},
  \citenamefont {Shishido}, \citenamefont {Hashimoto}, \citenamefont {Okubo},
  \citenamefont {Yamada}, \citenamefont {Inada}, \citenamefont {Settai},
  \citenamefont {Harima}, \citenamefont {Galatanu}, \citenamefont {Yamamoto},
  \citenamefont {Nakamura}, \citenamefont {Sugiyama}, \citenamefont {Takeuchi},
  \citenamefont {Kindo}, \citenamefont {Namiki}, \citenamefont {Aoki},
  \citenamefont {Sato},\ and\ \citenamefont {Ōnuki}}]{Ueda2004}%
  \BibitemOpen
  \bibfield  {author} {\bibinfo {author} {\bibfnamefont {T.}~\bibnamefont
  {Ueda}}, \bibinfo {author} {\bibfnamefont {H.}~\bibnamefont {Shishido}},
  \bibinfo {author} {\bibfnamefont {S.}~\bibnamefont {Hashimoto}}, \bibinfo
  {author} {\bibfnamefont {T.}~\bibnamefont {Okubo}}, \bibinfo {author}
  {\bibfnamefont {M.}~\bibnamefont {Yamada}}, \bibinfo {author} {\bibfnamefont
  {Y.}~\bibnamefont {Inada}}, \bibinfo {author} {\bibfnamefont
  {R.}~\bibnamefont {Settai}}, \bibinfo {author} {\bibfnamefont
  {H.}~\bibnamefont {Harima}}, \bibinfo {author} {\bibfnamefont
  {A.}~\bibnamefont {Galatanu}}, \bibinfo {author} {\bibfnamefont
  {E.}~\bibnamefont {Yamamoto}}, \bibinfo {author} {\bibfnamefont
  {N.}~\bibnamefont {Nakamura}}, \bibinfo {author} {\bibfnamefont
  {K.}~\bibnamefont {Sugiyama}}, \bibinfo {author} {\bibfnamefont
  {T.}~\bibnamefont {Takeuchi}}, \bibinfo {author} {\bibfnamefont
  {K.}~\bibnamefont {Kindo}}, \bibinfo {author} {\bibfnamefont
  {T.}~\bibnamefont {Namiki}}, \bibinfo {author} {\bibfnamefont
  {Y.}~\bibnamefont {Aoki}}, \bibinfo {author} {\bibfnamefont {H.}~\bibnamefont
  {Sato}},\ and\ \bibinfo {author} {\bibfnamefont {Y.}~\bibnamefont {Ōnuki}},\
  }\bibfield  {title} {\bibinfo {title} {{Electronic, Magnetic and
  Superconducting Properties of Quasi-two Dimensional Compounds Ce$_2$RhIn$_8$
  and La$_2$RhIn$_8$}},\ }\href {https://doi.org/10.1143/jpsj.73.649}
  {\bibfield  {journal} {\bibinfo  {journal} {J. Phys. Soc. Jpn.}\ }\textbf
  {\bibinfo {volume} {73}},\ \bibinfo {pages} {649} (\bibinfo {year}
  {2004})}\BibitemShut {NoStop}%
\bibitem [{\citenamefont {Kraftmakher}(2002)}]{Kraftmakher2002}%
  \BibitemOpen
  \bibfield  {author} {\bibinfo {author} {\bibfnamefont {Y.}~\bibnamefont
  {Kraftmakher}},\ }\bibfield  {title} {\bibinfo {title} {{Modulation
  calorimetry and related techniques}},\ }\href
  {https://doi.org/10.1016/s0370-1573(01)00031-x} {\bibfield  {journal}
  {\bibinfo  {journal} {Phys. Rep.}\ }\textbf {\bibinfo {volume} {356}},\
  \bibinfo {pages} {1} (\bibinfo {year} {2002})}\BibitemShut {NoStop}%
\bibitem [{\citenamefont {Michon}\ \emph {et~al.}(2019)\citenamefont {Michon},
  \citenamefont {Girod}, \citenamefont {Badoux}, \citenamefont {Kačmarčík},
  \citenamefont {Ma}, \citenamefont {Dragomir}, \citenamefont {Dabkowska},
  \citenamefont {Gaulin}, \citenamefont {Zhou}, \citenamefont {Pyon},
  \citenamefont {Takayama}, \citenamefont {Takagi}, \citenamefont {Verret},
  \citenamefont {Doiron-Leyraud}, \citenamefont {Marcenat}, \citenamefont
  {Taillefer},\ and\ \citenamefont {Klein}}]{Michon2019}%
  \BibitemOpen
  \bibfield  {author} {\bibinfo {author} {\bibfnamefont {B.}~\bibnamefont
  {Michon}}, \bibinfo {author} {\bibfnamefont {C.}~\bibnamefont {Girod}},
  \bibinfo {author} {\bibfnamefont {S.}~\bibnamefont {Badoux}}, \bibinfo
  {author} {\bibfnamefont {J.}~\bibnamefont {Kačmarčík}}, \bibinfo {author}
  {\bibfnamefont {Q.}~\bibnamefont {Ma}}, \bibinfo {author} {\bibfnamefont
  {M.}~\bibnamefont {Dragomir}}, \bibinfo {author} {\bibfnamefont {H.~A.}\
  \bibnamefont {Dabkowska}}, \bibinfo {author} {\bibfnamefont {B.~D.}\
  \bibnamefont {Gaulin}}, \bibinfo {author} {\bibfnamefont {J.-S.}\
  \bibnamefont {Zhou}}, \bibinfo {author} {\bibfnamefont {S.}~\bibnamefont
  {Pyon}}, \bibinfo {author} {\bibfnamefont {T.}~\bibnamefont {Takayama}},
  \bibinfo {author} {\bibfnamefont {H.}~\bibnamefont {Takagi}}, \bibinfo
  {author} {\bibfnamefont {S.}~\bibnamefont {Verret}}, \bibinfo {author}
  {\bibfnamefont {N.}~\bibnamefont {Doiron-Leyraud}}, \bibinfo {author}
  {\bibfnamefont {C.}~\bibnamefont {Marcenat}}, \bibinfo {author}
  {\bibfnamefont {L.}~\bibnamefont {Taillefer}},\ and\ \bibinfo {author}
  {\bibfnamefont {T.}~\bibnamefont {Klein}},\ }\bibfield  {title} {\bibinfo
  {title} {Thermodynamic signatures of quantum criticality in cuprate
  superconductors},\ }\href {https://doi.org/10.1038/s41586-019-0932-x}
  {\bibfield  {journal} {\bibinfo  {journal} {Nature}\ }\textbf {\bibinfo
  {volume} {567}},\ \bibinfo {pages} {218} (\bibinfo {year}
  {2019})}\BibitemShut {NoStop}%
\bibitem [{\citenamefont {Moshopoulou}\ \emph {et~al.}(2002)\citenamefont
  {Moshopoulou}, \citenamefont {Prokes}, \citenamefont {Garcia-Matres},
  \citenamefont {Pagliuso}, \citenamefont {Sarrao},\ and\ \citenamefont
  {Thompson}}]{Moshopoulou2002}%
  \BibitemOpen
  \bibfield  {author} {\bibinfo {author} {\bibfnamefont {E.}~\bibnamefont
  {Moshopoulou}}, \bibinfo {author} {\bibfnamefont {K.}~\bibnamefont {Prokes}},
  \bibinfo {author} {\bibfnamefont {E.}~\bibnamefont {Garcia-Matres}}, \bibinfo
  {author} {\bibfnamefont {P.}~\bibnamefont {Pagliuso}}, \bibinfo {author}
  {\bibfnamefont {J.}~\bibnamefont {Sarrao}},\ and\ \bibinfo {author}
  {\bibfnamefont {J.}~\bibnamefont {Thompson}},\ }\bibfield  {title} {\bibinfo
  {title} {{Neutron-diffraction study of field-induced transitions in the
  heavy-fermion compound Ce$_2$RhIn$_8$}},\ }\href
  {https://doi.org/10.1016/s0921-4526(02)00793-7} {\bibfield  {journal}
  {\bibinfo  {journal} {Physica B}\ }\textbf {\bibinfo {volume} {318}},\
  \bibinfo {pages} {300} (\bibinfo {year} {2002})}\BibitemShut {NoStop}%
\bibitem [{\citenamefont {Braithwaite}\ \emph {et~al.}(2019)\citenamefont
  {Braithwaite}, \citenamefont {Vališka}, \citenamefont {Knebel},
  \citenamefont {Lapertot}, \citenamefont {Brison}, \citenamefont {Pourret},
  \citenamefont {Zhitomirsky}, \citenamefont {Flouquet}, \citenamefont
  {Honda},\ and\ \citenamefont {Aoki}}]{Braithwaite2019}%
  \BibitemOpen
  \bibfield  {author} {\bibinfo {author} {\bibfnamefont {D.}~\bibnamefont
  {Braithwaite}}, \bibinfo {author} {\bibfnamefont {M.}~\bibnamefont
  {Vališka}}, \bibinfo {author} {\bibfnamefont {G.}~\bibnamefont {Knebel}},
  \bibinfo {author} {\bibfnamefont {G.}~\bibnamefont {Lapertot}}, \bibinfo
  {author} {\bibfnamefont {J.-P.}\ \bibnamefont {Brison}}, \bibinfo {author}
  {\bibfnamefont {A.}~\bibnamefont {Pourret}}, \bibinfo {author} {\bibfnamefont
  {M.~E.}\ \bibnamefont {Zhitomirsky}}, \bibinfo {author} {\bibfnamefont
  {J.}~\bibnamefont {Flouquet}}, \bibinfo {author} {\bibfnamefont
  {F.}~\bibnamefont {Honda}},\ and\ \bibinfo {author} {\bibfnamefont
  {D.}~\bibnamefont {Aoki}},\ }\bibfield  {title} {\bibinfo {title} {Multiple
  superconducting phases in a nearly ferromagnetic system},\ }\href
  {https://doi.org/10.1038/s42005-019-0248-z} {\bibfield  {journal} {\bibinfo
  {journal} {Commun. Phys.}\ }\textbf {\bibinfo {volume} {2}},\ \bibinfo
  {pages} {147} (\bibinfo {year} {2019})}\BibitemShut {NoStop}%
\bibitem [{\citenamefont {Yip}\ \emph {et~al.}(1991)\citenamefont {Yip},
  \citenamefont {Li},\ and\ \citenamefont {Kumar}}]{Yip1991}%
  \BibitemOpen
  \bibfield  {author} {\bibinfo {author} {\bibfnamefont {S.~K.}\ \bibnamefont
  {Yip}}, \bibinfo {author} {\bibfnamefont {T.}~\bibnamefont {Li}},\ and\
  \bibinfo {author} {\bibfnamefont {P.}~\bibnamefont {Kumar}},\ }\bibfield
  {title} {\bibinfo {title} {{Thermodynamic considerations and the phase
  diagram of superconducting UPt$_3$}},\ }\href
  {https://doi.org/10.1103/physrevb.43.2742} {\bibfield  {journal} {\bibinfo
  {journal} {Phys. Rev. B}\ }\textbf {\bibinfo {volume} {43}},\ \bibinfo
  {pages} {2742} (\bibinfo {year} {1991})}\BibitemShut {NoStop}%
\bibitem [{\citenamefont {Rosuel}\ \emph {et~al.}(2023)\citenamefont {Rosuel},
  \citenamefont {Marcenat}, \citenamefont {Knebel}, \citenamefont {Klein},
  \citenamefont {Pourret}, \citenamefont {Marquardt}, \citenamefont {Niu},
  \citenamefont {Rousseau}, \citenamefont {Demuer}, \citenamefont {Seyfarth},
  \citenamefont {Lapertot}, \citenamefont {Aoki}, \citenamefont {Braithwaite},
  \citenamefont {Flouquet},\ and\ \citenamefont {Brison}}]{Rosuel2023}%
  \BibitemOpen
  \bibfield  {author} {\bibinfo {author} {\bibfnamefont {A.}~\bibnamefont
  {Rosuel}}, \bibinfo {author} {\bibfnamefont {C.}~\bibnamefont {Marcenat}},
  \bibinfo {author} {\bibfnamefont {G.}~\bibnamefont {Knebel}}, \bibinfo
  {author} {\bibfnamefont {T.}~\bibnamefont {Klein}}, \bibinfo {author}
  {\bibfnamefont {A.}~\bibnamefont {Pourret}}, \bibinfo {author} {\bibfnamefont
  {N.}~\bibnamefont {Marquardt}}, \bibinfo {author} {\bibfnamefont
  {Q.}~\bibnamefont {Niu}}, \bibinfo {author} {\bibfnamefont {S.}~\bibnamefont
  {Rousseau}}, \bibinfo {author} {\bibfnamefont {A.}~\bibnamefont {Demuer}},
  \bibinfo {author} {\bibfnamefont {G.}~\bibnamefont {Seyfarth}}, \bibinfo
  {author} {\bibfnamefont {G.}~\bibnamefont {Lapertot}}, \bibinfo {author}
  {\bibfnamefont {D.}~\bibnamefont {Aoki}}, \bibinfo {author} {\bibfnamefont
  {D.}~\bibnamefont {Braithwaite}}, \bibinfo {author} {\bibfnamefont
  {J.}~\bibnamefont {Flouquet}},\ and\ \bibinfo {author} {\bibfnamefont
  {J.}~\bibnamefont {Brison}},\ }\bibfield  {title} {\bibinfo {title}
  {{Field-Induced Tuning of the Pairing State in a Superconductor}},\ }\href
  {https://doi.org/10.1103/physrevx.13.011022} {\bibfield  {journal} {\bibinfo
  {journal} {Phys. Rev. X}\ }\textbf {\bibinfo {volume} {13}},\ \bibinfo
  {pages} {011022} (\bibinfo {year} {2023})}\BibitemShut {NoStop}%
\bibitem [{\citenamefont {Bao}\ \emph {et~al.}(2000)\citenamefont {Bao},
  \citenamefont {Pagliuso}, \citenamefont {Sarrao}, \citenamefont {Thompson},
  \citenamefont {Fisk}, \citenamefont {Lynn},\ and\ \citenamefont
  {Erwin}}]{Bao2000}%
  \BibitemOpen
  \bibfield  {author} {\bibinfo {author} {\bibfnamefont {W.}~\bibnamefont
  {Bao}}, \bibinfo {author} {\bibfnamefont {P.~G.}\ \bibnamefont {Pagliuso}},
  \bibinfo {author} {\bibfnamefont {J.~L.}\ \bibnamefont {Sarrao}}, \bibinfo
  {author} {\bibfnamefont {J.~D.}\ \bibnamefont {Thompson}}, \bibinfo {author}
  {\bibfnamefont {Z.}~\bibnamefont {Fisk}}, \bibinfo {author} {\bibfnamefont
  {J.~W.}\ \bibnamefont {Lynn}},\ and\ \bibinfo {author} {\bibfnamefont
  {R.~W.}\ \bibnamefont {Erwin}},\ }\bibfield  {title} {\bibinfo {title}
  {{Incommensurate magnetic structure of CeRhIn$_5$}},\ }\href
  {https://doi.org/10.1103/physrevb.62.r14621} {\bibfield  {journal} {\bibinfo
  {journal} {Phys. Rev. B}\ }\textbf {\bibinfo {volume} {62}},\ \bibinfo
  {pages} {R14621} (\bibinfo {year} {2000})}\BibitemShut {NoStop}%
\bibitem [{\citenamefont {Correa}\ \emph {et~al.}(2005)\citenamefont {Correa},
  \citenamefont {Okraku}, \citenamefont {Betts}, \citenamefont {Migliori},
  \citenamefont {Sarrao},\ and\ \citenamefont {Lacerda}}]{Correa2005}%
  \BibitemOpen
  \bibfield  {author} {\bibinfo {author} {\bibfnamefont {V.~F.}\ \bibnamefont
  {Correa}}, \bibinfo {author} {\bibfnamefont {W.~E.}\ \bibnamefont {Okraku}},
  \bibinfo {author} {\bibfnamefont {J.~B.}\ \bibnamefont {Betts}}, \bibinfo
  {author} {\bibfnamefont {A.}~\bibnamefont {Migliori}}, \bibinfo {author}
  {\bibfnamefont {J.~L.}\ \bibnamefont {Sarrao}},\ and\ \bibinfo {author}
  {\bibfnamefont {A.~H.}\ \bibnamefont {Lacerda}},\ }\bibfield  {title}
  {\bibinfo {title} {{High-magnetic-field thermal expansion and elastic
  properties of CeRhIn$_5$}},\ }\href
  {https://doi.org/10.1103/physrevb.72.012407} {\bibfield  {journal} {\bibinfo
  {journal} {Phys. Rev. B}\ }\textbf {\bibinfo {volume} {72}},\ \bibinfo
  {pages} {012407} (\bibinfo {year} {2005})}\BibitemShut {NoStop}%
\bibitem [{\citenamefont {Raymond}\ \emph {et~al.}(2007)\citenamefont
  {Raymond}, \citenamefont {Ressouche}, \citenamefont {Knebel}, \citenamefont
  {Aoki},\ and\ \citenamefont {Flouquet}}]{Raymond2007}%
  \BibitemOpen
  \bibfield  {author} {\bibinfo {author} {\bibfnamefont {S.}~\bibnamefont
  {Raymond}}, \bibinfo {author} {\bibfnamefont {E.}~\bibnamefont {Ressouche}},
  \bibinfo {author} {\bibfnamefont {G.}~\bibnamefont {Knebel}}, \bibinfo
  {author} {\bibfnamefont {D.}~\bibnamefont {Aoki}},\ and\ \bibinfo {author}
  {\bibfnamefont {J.}~\bibnamefont {Flouquet}},\ }\bibfield  {title} {\bibinfo
  {title} {{Magnetic structure of CeRhIn$_5$ under magnetic field}},\ }\href
  {https://doi.org/10.1088/0953-8984/19/24/242204} {\bibfield  {journal}
  {\bibinfo  {journal} {J. Phys. Condens. Matter}\ }\textbf {\bibinfo {volume}
  {19}},\ \bibinfo {pages} {242204} (\bibinfo {year} {2007})}\BibitemShut
  {NoStop}%
\bibitem [{\citenamefont {Fobes}\ \emph {et~al.}(2018)\citenamefont {Fobes},
  \citenamefont {Zhang}, \citenamefont {Lin}, \citenamefont {Das},
  \citenamefont {Ghimire}, \citenamefont {Bauer}, \citenamefont {Thompson},
  \citenamefont {Harriger}, \citenamefont {Ehlers}, \citenamefont {Podlesnyak},
  \citenamefont {Bewley}, \citenamefont {Sazonov}, \citenamefont {Hutanu},
  \citenamefont {Ronning}, \citenamefont {Batista},\ and\ \citenamefont
  {Janoschek}}]{Fobes2018}%
  \BibitemOpen
  \bibfield  {author} {\bibinfo {author} {\bibfnamefont {D.~M.}\ \bibnamefont
  {Fobes}}, \bibinfo {author} {\bibfnamefont {S.}~\bibnamefont {Zhang}},
  \bibinfo {author} {\bibfnamefont {S.-Z.}\ \bibnamefont {Lin}}, \bibinfo
  {author} {\bibfnamefont {P.}~\bibnamefont {Das}}, \bibinfo {author}
  {\bibfnamefont {N.~J.}\ \bibnamefont {Ghimire}}, \bibinfo {author}
  {\bibfnamefont {E.~D.}\ \bibnamefont {Bauer}}, \bibinfo {author}
  {\bibfnamefont {J.~D.}\ \bibnamefont {Thompson}}, \bibinfo {author}
  {\bibfnamefont {L.~W.}\ \bibnamefont {Harriger}}, \bibinfo {author}
  {\bibfnamefont {G.}~\bibnamefont {Ehlers}}, \bibinfo {author} {\bibfnamefont
  {A.}~\bibnamefont {Podlesnyak}}, \bibinfo {author} {\bibfnamefont {R.~I.}\
  \bibnamefont {Bewley}}, \bibinfo {author} {\bibfnamefont {A.}~\bibnamefont
  {Sazonov}}, \bibinfo {author} {\bibfnamefont {V.}~\bibnamefont {Hutanu}},
  \bibinfo {author} {\bibfnamefont {F.}~\bibnamefont {Ronning}}, \bibinfo
  {author} {\bibfnamefont {C.~D.}\ \bibnamefont {Batista}},\ and\ \bibinfo
  {author} {\bibfnamefont {M.}~\bibnamefont {Janoschek}},\ }\bibfield  {title}
  {\bibinfo {title} {Tunable emergent heterostructures in a prototypical
  correlated metal},\ }\href {https://doi.org/10.1038/s41567-018-0060-9}
  {\bibfield  {journal} {\bibinfo  {journal} {Nat. Phys.}\ }\textbf {\bibinfo
  {volume} {14}},\ \bibinfo {pages} {456} (\bibinfo {year} {2018})}\BibitemShut
  {NoStop}%
\bibitem [{\citenamefont {Honda}\ \emph {et~al.}(2014)\citenamefont {Honda},
  \citenamefont {Taga}, \citenamefont {Hirose}, \citenamefont {Yoshiuchi},
  \citenamefont {Tomooka}, \citenamefont {Ohya}, \citenamefont {Sakaguchi},
  \citenamefont {Takeuchi}, \citenamefont {Settai}, \citenamefont {Shimura},
  \citenamefont {Sakakibara}, \citenamefont {Sheikin}, \citenamefont {Tanaka},
  \citenamefont {Kubo},\ and\ \citenamefont {Ōnuki}}]{Honda2014}%
  \BibitemOpen
  \bibfield  {author} {\bibinfo {author} {\bibfnamefont {F.}~\bibnamefont
  {Honda}}, \bibinfo {author} {\bibfnamefont {Y.}~\bibnamefont {Taga}},
  \bibinfo {author} {\bibfnamefont {Y.}~\bibnamefont {Hirose}}, \bibinfo
  {author} {\bibfnamefont {S.}~\bibnamefont {Yoshiuchi}}, \bibinfo {author}
  {\bibfnamefont {Y.}~\bibnamefont {Tomooka}}, \bibinfo {author} {\bibfnamefont
  {M.}~\bibnamefont {Ohya}}, \bibinfo {author} {\bibfnamefont {J.}~\bibnamefont
  {Sakaguchi}}, \bibinfo {author} {\bibfnamefont {T.}~\bibnamefont {Takeuchi}},
  \bibinfo {author} {\bibfnamefont {R.}~\bibnamefont {Settai}}, \bibinfo
  {author} {\bibfnamefont {Y.}~\bibnamefont {Shimura}}, \bibinfo {author}
  {\bibfnamefont {T.}~\bibnamefont {Sakakibara}}, \bibinfo {author}
  {\bibfnamefont {I.}~\bibnamefont {Sheikin}}, \bibinfo {author} {\bibfnamefont
  {T.}~\bibnamefont {Tanaka}}, \bibinfo {author} {\bibfnamefont
  {Y.}~\bibnamefont {Kubo}},\ and\ \bibinfo {author} {\bibfnamefont
  {Y.}~\bibnamefont {Ōnuki}},\ }\bibfield  {title} {\bibinfo {title} {{Novel
  Electronic States of Heavy Fermion Compound YbCo$_2$Zn$_{20}$}},\ }\href
  {https://doi.org/10.7566/jpsj.83.044703} {\bibfield  {journal} {\bibinfo
  {journal} {J. Phys. Soc. Jpn.}\ }\textbf {\bibinfo {volume} {83}},\ \bibinfo
  {pages} {044703} (\bibinfo {year} {2014})}\BibitemShut {NoStop}%
\bibitem [{\citenamefont {Tayama}\ \emph {et~al.}(2003)\citenamefont {Tayama},
  \citenamefont {Sakakibara}, \citenamefont {Sugawara}, \citenamefont {Aoki},\
  and\ \citenamefont {Sato}}]{Tayama2003}%
  \BibitemOpen
  \bibfield  {author} {\bibinfo {author} {\bibfnamefont {T.}~\bibnamefont
  {Tayama}}, \bibinfo {author} {\bibfnamefont {T.}~\bibnamefont {Sakakibara}},
  \bibinfo {author} {\bibfnamefont {H.}~\bibnamefont {Sugawara}}, \bibinfo
  {author} {\bibfnamefont {Y.}~\bibnamefont {Aoki}},\ and\ \bibinfo {author}
  {\bibfnamefont {H.}~\bibnamefont {Sato}},\ }\bibfield  {title} {\bibinfo
  {title} {{Magnetic Phase Diagram of the Heavy Fermion Superconductor
  PrOs$_4$Sb$_{12}$}},\ }\href {https://doi.org/10.1143/jpsj.72.1516}
  {\bibfield  {journal} {\bibinfo  {journal} {J. Phys. Soc. Jpn.}\ }\textbf
  {\bibinfo {volume} {72}},\ \bibinfo {pages} {1516} (\bibinfo {year}
  {2003})}\BibitemShut {NoStop}%
\end{thebibliography}%

\end{document}